\documentclass[aps,pra,twocolumn,nofootinbib,10pt]{revtex4-2}

\usepackage{cancel}
\usepackage{xcolor}
\usepackage{footmisc}
\usepackage{subcaption}
\usepackage{footnote}
\makesavenoteenv{figure}
\usepackage{ragged2e}
\usepackage{amsmath, amssymb, booktabs}
\usepackage{graphicx}% Include figure files
\usepackage{dcolumn}% Align table columns on decimal point
\usepackage{bm}% bold math
\usepackage{physics}
\usepackage{hyperref}
\usepackage{comment}

\usepackage{tikz}
\usetikzlibrary{shapes.geometric, arrows}
\tikzstyle{process} = [rectangle,
minimum width=8cm,
minimum height=1.1cm,
text centered,
text width=8cm,
draw=black,
fill=red!30]
\tikzstyle{arrow} = [thick,->,>=stealth]
\begin{document}

\preprint{APS/123-QED}

\title{Random acceleration noise on Stern-Gerlach Interferometry in a Harmonic Trap} 

\author{Sneha Narasimha Moorthy$^{1, \,2}$}
\author{Andrew Geraci $^{3}$}
%\author{Tracy Northup $^{4}$}
\author{Sougato Bose $^{4}$}
\author{Anupam Mazumdar$^{5}$}
\affiliation{ 
$^{1}$School of Physical Sciences, National Institute of Science Education and Research, Jatni 752050, India\\
$^{2}$Homi Bhabha National Institute, Training School Complex, Anushaktinagar, Mumbai 400094, India\\
$^{3}$ Center for Fundamental Physics, Department of Physics and
Astronomy, Northwestern University, 2145 Sheridan Road, Evanston, IL, USA\\
%$^{4}$ Dept. of Experimental Physics, University of Innsbruck, Technikerstraße 25, 6020, Innsbruck, Austria\\
$^{4}$ Department of Physics and Astronomy, University College London, Gower Street, WC1E 6BT London, United Kingdom\\
$^{5}$ Van Swinderen Institute, University of Groningen, 9747 AG Groningen, The Netherlands\\}

\date{\today}% It is always \today, today,

\begin{abstract}
We analyze decoherence in a one-loop Stern--Gerlach--type matter-wave interferometer for a massive nanoparticle embedded with a nitrogen vacancy (NV)-centered nanodiamond evolving under an effective harmonic-oscillator dynamics in a magnetic-field gradient. We assume that the Stern-Gerlach interferometer is subjected to an acceleration $\vec{a}$ external to the system, which is at an angle $\theta_0$ with respect to the direction of the superposition.
For a one loop interferometer, we quantify dephasing from two noise channels: fluctuations in the external acceleration $\delta a(t)$ and fluctuations in the tilt angle $\delta \theta(t)$. At the level of the action, we treat these two external noise as stochastic inputs, compute the resulting stochastic phase difference between the interferometer arms, and obtain the dephasing rate $\Gamma$. We obtain the transfer function of the interferometer for each of the noise sources. We also show explicit results of constraints on the power spectral density of the noise sources for an interferometer that produces a superposition size of $\Delta x\sim 1$nm of a nanodiamond of mass $m=10^{-15}~\mathrm{kg}$ by considering white noise statistics and imposing a coherence target $\Gamma \tau\leq 1$, where $\tau\simeq 0.015~\mathrm{s}$. We find $\sqrt{\mathcal{S}_{aa}}\lesssim \mathcal{O}(10^{-11})~\mathrm{m\,s^{-2}\,Hz^{-1/2}}$ if we take the external acceleration, $a=0~{\rm ms^{-2}}$ and $\theta_0=0^\circ$ (along the direction of the superposition), and $\sqrt{\mathcal{S}_{\theta\theta}}\lesssim \mathcal{O}(10^{-10})~\mathrm{rad\,Hz^{-1/2}}$ for $a=g= 9.81~\mathrm{m\,s^{-2}}$ and $\theta_0=90^\circ$ (superposition direction is perpendicular to the Earth's gravity). We have also found an operating regime where the acceleration noise can be minimized by either varying $\theta_0$ or $a$ for a fixed set of other experimental parameters.
\end{abstract}

\maketitle

%\tableofcontents

\section{Introduction}

Matter-wave interferometers are becoming increasingly sophisticated to test fundamental physics involving quantum-mechanical principles for heavy masses~\cite{Colella:1975dq,Werner:1979gi,Rauch:2015jkh}, including the large spatial superposition~\cite{Fixler:2007is,Asenbaum:2016djh,Overstreet:2021hea,Amit_2019,SGI_experiment,Kovachi_2015}, and states of a massive object~\cite{Arndt_1999,Eibenberger:2013cqb,Fein:2019dgf,pedalino2025,arndt2014testing}. By increasing the mass, we become more sensitive to the gravitational interaction, leading to a recent protocol to test the quantum nature of the gravitational interaction with matter by keeping two adjacent matter-wave interferometers and studying quantum features such as entanglement to witness the quantum nature of gravity, known as quantum gravity induced entanglement of matter (QGEM) \cite{Bose:2017nin,ICTS,marletto2017gravitationally}. Similar to the light-bending experiment due to gravity, one proposed test is to examine how the quantum-gravitational interaction can lead to entanglement between matter and a photon via the exchange of a virtual graviton~\cite{Biswas:2022qto}. Similarly, with the help of matter wave interferometers, one can probe the Kaluza-Klein theory of massive graviton and entanglement~\cite {Elahi:2024dbb}, non-local gravitational interactions~\cite{Vinckers:2023grv}, and test the weak equivalence principle via witnessing entanglement~\cite{chakraborty2023distinguishing,Bose:2022czr}.

There are many challenges to overcome before we can reach the precision to create a large macroscopic quantum superposition. There are many sources of decoherence for massive quantum systems~\cite{bassireview,RomeroIsart2011LargeQS} due to random external fluctuations in ambient temperature/pressure~\cite{RomeroIsart2011LargeQS,vandeKamp:2020rqh,Schut:2021svd,Schut:2024lgp,Rijavec:2020qxd}, systematics related to initialization \cite{Schut:2023eux}, current/magnetic field fluctuations~\cite{Sneha_HP,Sneha_IHP}, external dipoles~\cite{Fragolino:2023agd}, voltage~\cite{Leibfried:2003zz}, random accelerations and gravity gradients ~\cite{RelAcc, Wu:2024bzd}. There are phonon-induced noise~\cite{Henkel:2021wmj,Henkel:2023tqe,Xiang:2024zol}, and fluctuation in the spin degrees of freedom during the dynamics of rotation of the rigid body~\cite{Stickler:2021dho,Stickler:2018uii,Rusconi:2022jhm,
Japha:2022phw,Zhou:2024pdl,Rizaldy:2024viw}, all leading to dephasing and decoherence, see~\cite{bassireview,ORI11_GM,Hornberger_2012}, and loss of contrast~\cite{Englert,Schwinger,Scully,SGI_experiment}.

One particularly popular platform for creating large spatial superposition is within diamagneticaly levitated nanoparticle with a defect such as the nitrogen vacancy (NV)-centred nanodiamond, see for a review on NV properties of nanodiamond~\cite{Doherty_2013}. For NV-centered nanodiamonds, it is possible to create a spin superposition of 
$|+1\rangle$ and $|-1\rangle$ states. It is also possible to cool a nanoparticle to a motional ground state in optical and in ion traps, see~\cite{Deli__2020,Piotrowski_2023,Kamba:2023zoq,Bykov:2022xji,Perdriat:2024xiy,Bykov:2022xji}, and nanodaimonds in a diamagnetic traps, see~\cite{Hsu:2016}. Once motionally cooled the idea is to apply the Stern-Gerlach inhomogeneous magnetic field on the spin of NV~\cite{SGI_experiment,Amit_2019} to separate the centre of mass motion of the nanodiamond onto left and right trajectories, before bringing them together to perform one-loop interferometer, thereby creating a Schr\"odinger cat state. 

The current paper analyses a small part of the whole lot of noise budget, but a significant one from the experimental perspective. The current analysis focuses on characterizing the dephasing\footnote{{ We use the terms decoherence and dephasing interchangeably in this paper - we are referring to the phase accumulated due stochastic noise in the environment that enters the phase measurement and affects the subsequent averaging over experimental runs.}} introduced by noise in an external acceleration field, like external jitters/vibrations in the vicinity of the experiment, by modelling it using classical stochastic processes at the Lagrangian level. We consider the impact of Gaussian white noise on the phase evolution of the interferometric superposition. Additionally, we analyze the dephasing due to fluctuations in the tilt angle between the superposition axis and the external acceleration field. We obtain bounds on the noise parameters, including the power spectral density (PSD) describing the acceleration and tilt fluctuation statistics.{ We would like to emphasize that the acceleration and the tilt-angle fluctuations need not have the same physical origin and hence need not be statistically correlated. In special experimental situations the same source of noise may generate both acceleration and tilt fluctuations, in which case their cross-correlations should be included. In this work, however, we treat acceleration noise and tilt-angle noise as separate limiting cases.} We also note that, unlike the tilt fluctuations~\cite{Schut:2023eux} and magnetic noise fluctuations, \cite{Sneha_HP,Sneha_IHP}, {which are associated with controllable experimental parameters like alignment and magnetic-field stability}, the gravitational field fluctuation is a true environmental fluctuation, whose origin can be seismic~\cite{Hughes_1998}, or a random acceleration noise~\cite{RelAcc}, similar in the context of gravitational wave detectors~\cite{Saulson:1984,Thorne_1999},see for a review~\cite{Harms_2019}.

{Our analysis is complementary to Ref.~\cite{Wu:2024tcr}, which studied dephasing from a broader class of noises and emphasized the general linear-response structure of the problem. The goal of the present work is different. We specialize to acceleration and tilt-angle fluctuations relevant to a Stern--Gerlach nanoparticle interferometer and translate the corresponding phase-noise transfer functions into quantitative constraints on experimentally tolerable noise levels. In particular, we do not assume that acceleration fluctuations are necessarily common-mode between the two interferometer arms. If the fluctuating source is local, or if the spatial correlation length of the noise is comparable to or smaller than the arm separation, common-mode rejection is reduced. The uncorrelated-arm treatment considered here therefore gives a useful non-common-mode benchmark for experimental design, while the fully correlated case discussed in Ref.~\cite{Wu:2024tcr} provides the opposite limiting case.

The central output of this paper is to derive the transfer function for acceleration noise and tilt angle noise, independently. We then use the transfer function to convert run-to-run variations and stochastic fluctuations into a quantitative dephasing rate. This is then translated into bounds on the tolerable noise power spectral density for each of the acceleration and tilt-angle fluctuations for representative Stern--Gerlach nanoparticle-interferometer parameters. We also identify parameter regimes where the interferometer is less susceptible to acceleration noise and compare the resulting requirements with experimentally relevant acceleration and tilt-noise scales.}

%{\bf AM: we should bring in Ref.[71] here and head on say the differences between this paper and that of the Ref.[71], here we should also emphasise why this paper is important for experimentalists. }

\section{Setup}

We consider a massive nanodiamond of mass $m\sim10^{-15}$~kg with a single nitrogen--vacancy (NV) center, whose ground-state electronic spin is \(S=1\)~\footnote{The choice of the mass is based on the QGEM experiment, see~\cite{Schut:2023hsy,Schut:2025blz}. We can vary the mass in our analysis.}.
We prepare the NV spin in an equal superposition of the \(\lvert m_x=+1\rangle\) and \(\lvert m_x=-1\rangle\) eigenstates of \(S_x\), and use a spin-dependent force to spatially separate the nanoparticle's centre-of-mass wavepacket into two branches, thereby entangling spin and position degrees of freedom. We shall focus only on the dynamics along the x-axis (one-dimensional matter-wave interferometer), along which spatial superposition is intended to be created. 
{ In the following, the spin is aligned by the external magnetic field, and we assume that the bias magnetic field will suppress spin flips during the interferometer sequence. Further, the phase of each branch is computed from the classical action along the corresponding centre-of-mass trajectory. This is the usual semiclassical approximation used in Stern--Gerlach-type interferometers.}

We consider a general case when the axis of spatial separation makes an arbitrary angle with the direction of acceleration due to some external force, like vibration of a table and gravitational acceleration. The corresponding Lagrangian for each of the arms ($j\in \{R,L\}$) of the interferometer is~\cite{Pedernales:2020nmf,Marshman:2021wyk}:
\begin{align}
    L_j &= \frac{1}{2}m\dot{x}_j^2 +\frac{\chi_\rho m}{2\mu_0}B_{xj}^2 - \hbar\gamma_e S_{xj} B_{xj} \nonumber \\
    &\quad- \hbar DS_{NV}^2 + max_j\cos\theta_0 \label{eq.EffPSD_genLag}
\end{align}
%{\bf AM: here you have used the subscript $NV$, but in the rest of the paper you omitted it, so better keep it uniform and remove $NV$, and on top of that, you have $S_x$, actually, these two are the same.}

where $R$ and $L$ refer to the right and left arms of the interferometer, and $S_{xR} = 1$ and $S_{xL} = -1$. The first term is the kinetic term, the second term is the diamagnetic induced contribution, where $\chi_{\rho}= -6.286\times10^{-9} \text{ m}^3 \text{ kg}^{-1}$ (for nanodiamond) represents the mass magnetic susceptibility, ~$\mu_0=4\pi\times 10^{-7}\text{ H m}^{-1}$ is the magnetic permeability, $\hbar=1.05\times 10^{-34}\text{ kg m}^2\text{ s}^{-1}$ is the reduced Planck's constant, $\gamma_e= 1.761 \times 10^{11}\text{ s}^{-1} \, \text{ T}^{-1}$ is the electron gyromagnetic ratio. The external magnetic field at $x_j$ is given by $B_{xj}$~\footnote{The dependence of $B_{xj}$ on the interferometer arm accounts for non-uniform magnetic fields of the form: $B_x\simeq B_0+\eta_1 x$.}. The third term indicates the spin-magnetic field coupling, $\vec{\mu}_S\cdot\vec{B}$ where $\vec{\mu}$ is the spin magnetic moment~\footnote{$\vec{S}\cdot \vec{B}$ term gives rise to a spin-dependent potential. This allows for entangling spin and position degrees of freedom, thus creating the two arms of the interferometer.}. The parameter \( D \) denotes the zero-field splitting, which for NV centers is \( D = {{2\pi\times}2.87 \, \mathrm{GHz}} \) \cite{Gruber2012-1997}. {$S_{NV}$ is the spin along the NV axis of the NV-centered; it is taken to be along the direction of creating the spatial superposition which in our case is the $x$-axis\footnote{The orientation of the NV axis and the biased magnetic field are taken to be along the interferometer axis to avoid precession. This allows us to assume 1D dynamics for the present analysis.}.} The last term in the Lagrangian is introduced by the axis of spatial superposition, making an angle $\theta_0$ with respect to the direction of acceleration $\vec{a} = (a_x, a_y, a_z) \equiv a\hat{a}$. In other words, since we have considered the interferometer axis along $x$, i.e. $ a_x = a\cos\theta_0 = \hat{x}\cdot\vec{a}$.

To study the effect of fluctuations in the gravitational acceleration, we shall use a simple model of the interferometer setup, assuming the magnetic field to be $B_{x} = B_0 + \eta_0 x$, where $B_0$ is the biased magnetic field. Thus, the Lagrangian becomes:
\begin{align}
    L_j = & \frac{1}{2} m \dot{x}_j^2 - \frac{1}{2} m \omega_0^2 x_j^2 
    - C_j\eta_0x_j + \frac{\chi_\rho m}{2 \mu_0} B_0^2 \nonumber \\
    & - S_{xj}\hbar \gamma_e B_0 - \hbar D S_{NV}^2+ max_j\cos\theta_0, \label{eq.Lag_j_HP}
\end{align}
where, 
\begin{equation}
    C_j = \left(S_{xj}\hbar\gamma_e - \frac{ \chi_\rho m}{\mu_0}B_{0}\right) \label{eq.Cj_def}
\end{equation}
and the characteristic frequency \( \omega_0 \) is defined as:
\begin{align}
    \omega_0 = \left( -\frac{\chi_\rho}{\mu_0} \right)^{1/2} \eta_0. \label{eq.omega0_def}
\end{align}

%%%%%%%%%%%%%%%%%%%%%%%%%%%%%%%%%%%%%%%%%%%
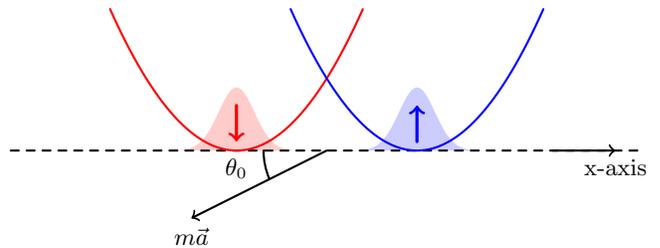
\begin{figure}[t]
\centering\begin{tikzpicture}[scale=1.2, line cap=round, line join=round]

% ---------- Axes ----------
\draw[thick, black, dashed] (-3.5,0) -- (3.5,0);    
\draw[thick, black, ->] (2.5,0) -- (3.2,0)node[below] {x-axis};
% horizontal reference
\def\ang{23} % arctan(1.5/3)
\draw[thick, black, ->] (0,0) -- (-1.5,-0.75) node[below] {$m\vec{a}$};

% ---------- Angle theta ----------
\draw[black, thick] (-0.7,0) arc[start angle=180, end angle=180+\ang, radius=0.8];
\node at (-1,-0.2) {$\theta_0$};

% Convenience: unit vector along the tilted guide direction
% (TikZ rotates in degrees; use \ang)
\coordinate (O) at (0,0);

% ---------- Rotated "up" and "down" arrows along the guide ----------
% "Up" along +guide direction

% "Down" along -guide direction

% ---------- Parameters for bell curve ----------
\def\A{0.7}      % amplitude (visual height)
\def\sigma{0.2}  % width parameter

% ---------- Left trap (tilted) ----------
\begin{scope}[shift={(-1,0)}, rotate=0]
  % Harmonic potential
  \draw[red, thick, domain=-1.4:1.4, samples=140]
    plot (\x,{0.8*(\x)^2});

  % Bell curve (Gaussian) centered at the minimum
  \fill[red, opacity=0.2, domain=-1.0:1.0, samples=160]
  plot (\x,{\A*exp(-(\x*\x)/(2*\sigma*\sigma))})
  -- (1.0,0) -- (-1.0,0) -- cycle;
\end{scope}
\begin{scope}[shift={(-1,0)}, rotate=90]
  \draw[very thick, red, ->] (0.5,0) -- (0.1,0) ;
\end{scope}

% ---------- Right trap (tilted) ----------
\begin{scope}[shift={(1,0)}, rotate=0]
  % Harmonic potential
  \draw[blue, thick, domain=-1.4:1.4, samples=140]
    plot (\x,{0.8*(\x)^2});

  % Bell curve (Gaussian)
  \fill[blue, opacity=0.2, domain=-1.0:1.0, samples=160]
  plot (\x,{\A*exp(-(\x*\x)/(2*\sigma*\sigma))})
  -- (1.0,0) -- (-1.0,0) -- cycle;
\end{scope}

\begin{scope}[shift={(1,0)}, rotate=90]
  \draw[very thick, blue, ->] (0.1,0) -- (0.5,0);
\end{scope}

\end{tikzpicture}

\caption{
Schematic of a nanoparticle interferometer in a tilted spin-dependent harmonic potential.
The magnetic-field gradient produces spin-dependent harmonic oscillator potentials.
External random acceleration is acting at an angle $\theta_0$ to the axis of spatial superposition. The gaussian wavepackets represent the states constituting the two interferometer arms.}
\label{Fig:0}
\end{figure}
For a one-loop interferometer, the time for a single run of the experiment would be: $\tau={2\pi}/{\omega_0}$. In the analysis of the current paper, we consider a one-loop interferometer. The equation of motion is as follows:
\begin{align}
    m\ddot{x}_j(t)&= -m\omega_0^2x_j(t) - C_j\eta_0 + ma\cos\theta_0\label{eq.EOM_det}
\end{align}
Imposing $x_j(0) = 0$ and $\dot{x}(0) = 0$, we get:
\begin{equation}
    x_j(t) = \frac{C_j \eta_0-ma\cos\theta_0}{m\omega_0^2 } (\cos(\omega_0 t) - 1) 
\end{equation}
We define:
\begin{align}
    \tilde{C}_j = C_j \eta_0-ma\cos\theta_0 \label{eq.tilde_C_j}
\end{align}
then the trajectory of each arm of the interferometer is given by:
\begin{equation}
    x_j(t) = \frac{\tilde{C}_j}{m\omega_0^2 } (\cos(\omega_0 t) - 1) \label{eq.det_traj}
\end{equation}

\begin{figure}[ht!]
    \centering
    \includegraphics[width=0.9\linewidth]{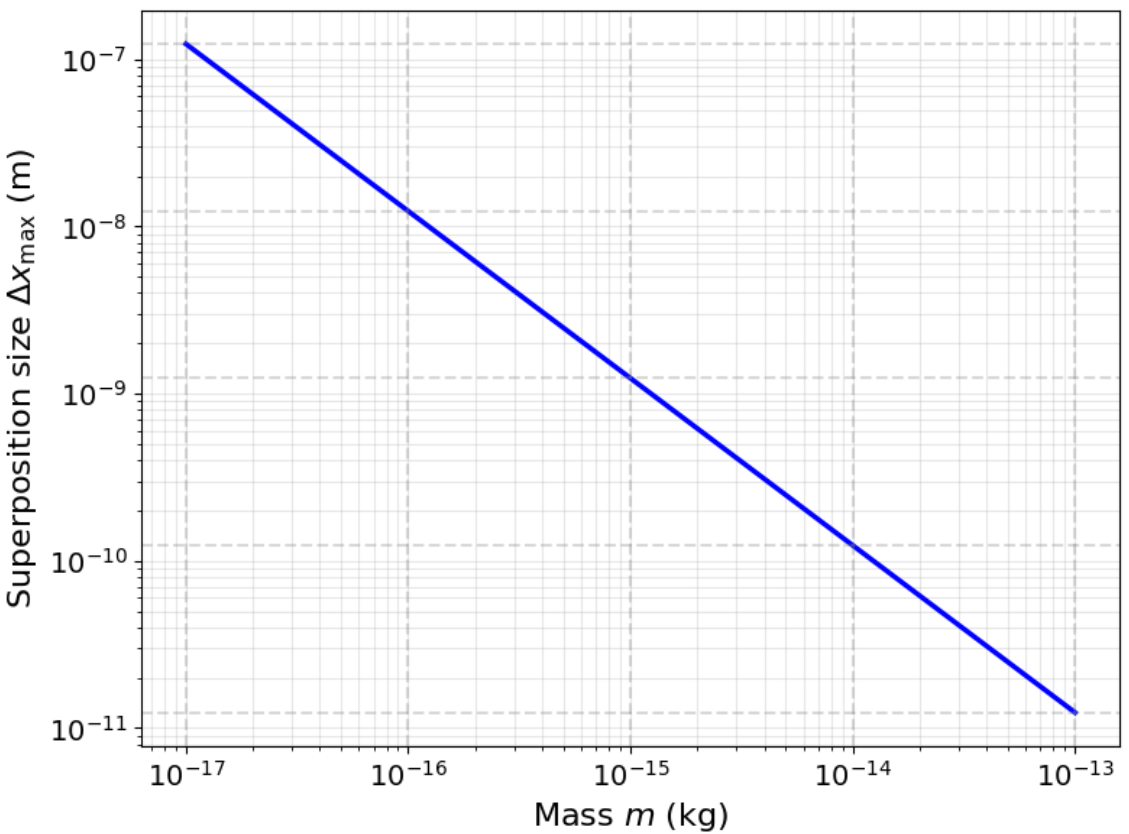}
    \caption{The graph depicts the variation of the maximum superposition size achievable in a harmonic potential as a function of the mass of the nanoparticle in a Stern-Gerlach-type interferometer for a set of parameters given by: $~\eta_0=6\times 10^{3}{\rm \  T\ m^{-1}}$, $m=10^{-15}{\rm kg}$.} %{\bf AM:  It is not clear to me what you meant here. You are varying m in x-axis.}
    \label{fig:superposition_size_vs_m}
\end{figure}

The maximum spatial superposition size ($\Delta x_\text{max}$) is obtained for $t= \frac{\pi}{\omega_0}$:
\begin{align}
    \Delta x_\text{max} &= \bigg|x_R\left(\frac{\pi}{\omega_0}\right) - x_L\left(\frac{\pi}{\omega_0}\right)\bigg| = 2\frac{\tilde{C}_R -\tilde{C}_L}{m\omega_0^2 }\nonumber\\
    &=4\frac{\hbar\gamma_e\eta_0}{m\omega_0^2 }
\end{align}
Hence, the superposition size doesn't depend on the magnetic field offset $B_0$ or on the external acceleration parameters $a$ and $\theta_0$. However, the superposition size does depend on mass as shown in Fig.\ref{fig:superposition_size_vs_m}. Note that the mass will also affect the dephasing parameters in both $\theta$ and acceleration.

%%%%%%%%%%%%%%%%%%%%%%%%%%%%
%\begin{table}[ht!]
%\centering
%\begin{tabular}{|W{c}{0.2\linewidth}|W{c}{0.4\linewidth}|}
%\hline
%Parameter & Value \\
%\hline
%$a$ & $g \simeq 9.81~\mathrm{m\,s^{-2}}$ 
%\\
%$\theta_0$ & $90^\circ$ 
% \\
%$\eta_0$ & $6\times 10^3~\mathrm{T\,m^{-1}}$ 
%\\
%$m$ & $10^{-15}~\mathrm{kg}$ 
%\\
%$\tau$ & $T=\dfrac{2\pi}{\omega_0}\simeq %0.015~\mathrm{s}$ 
%\\
%\hline
%\end{tabular}
%\caption{These are the parameter values we have used to evaluate the acceleration-noise, angular noise and the superposition size. }\label{tab:params_acc_noise}
%\end{table}

%%%%%%%%%%%%%%%%%%%%%%%%%%%%%
To study the effect of fluctuations in the acceleration $a$ on dephasing the interferometer, we briefly recap noise analysis in the following section. 

%%%%%%%%%%%%%%%%%%%%%%%%

\section{Noise Analysis}

We first consider a general Lagrangian of the form:
\begin{equation}
    \mathcal{L}_j = \frac{1}{2}m v_j^2 - A_j x_j^2 - B_j x_j - C_j, \;\;\text{where}\; j \in \{L, R\}.
\end{equation}
and $j$ represents the right and left arms of the interferometer. We consider a noise source, $\delta \zeta(t)$, which leads to fluctuations in the coefficients of the Lagrangian, for instance $A_j\to A_j+\delta A_j(t)$ if $A_j$ depends on $\zeta$. The modified Lagrangian gives rise to modified equations of motion and hence modified trajectories, say $x_j\to x_j+\delta x_j$. If the statistics of the noise are known, we can compute the dephasing rate using the following method, see for instance ~\cite{Sneha_HP, Sneha_IHP}:

%in the flowchart given in Fig.\ref{fig:flowchart_noise_analysis}.

\begin{itemize}
    \item Compute the relative phase between the two arms of the interferometer \cite{Storey1994PathIntegralAtomInterferometry,FeynmanHibbs1965}: $\Delta \phi = \frac{1}{\hbar} \int_{t_i}^{t_f} (\mathcal{L}_R - \mathcal{L}_L) \, dt$.
    \item Identify noise sources affecting coefficients $A_j$, $B_j$, $C_j$, introducing time-dependent fluctuations $\delta A_j(t)$, $\delta B_j(t)$, $\delta C_j(t)$. Characterise their statistical properties (generally in the frequency space, including power spectral density $\mathcal{S}(\omega)$).
    \item Derive equations of motion (EOM) for trajectories $x_j$ (noiseless) and $x_j^{\text{tot}}$ (with noise). Compute trajectory deviation: $\delta x_j = x_j^{\text{tot}} - x_j$.
    \item Calculate the phase fluctuation due to noise\footnote{{In Eq.\eqref{eq.general phase fluctuation}, $\delta x_j^2$ is the fluctuation in $x_j^2$ not the square of the first order fluctuation.}}: 
    {\small\begin{align}
        \delta \phi &= \frac{1}{\hbar} \int_{t_i}^{t_f} \bigg[ (\delta A_R x_R^2 - \delta A_L x_L^2)  \nonumber\\
        &\quad\quad\quad+ (\delta B_R x_R - \delta B_L x_L)+ (\delta C_R - \delta C_L) \bigg] \, dt \nonumber\\
        &\quad+ \frac{1}{\hbar} \int_{t_i}^{t_f}\bigg[ ( A_R \delta x_R^2 -  A_L \delta x_L^2)\nonumber\\
        &\quad\quad\quad\quad + ( B_R \delta x_R -  B_L \delta x_L) \bigg]  \, dt\label{eq.general phase fluctuation}
    \end{align}}
    \item Evaluate the dephasing rate, see~\cite{Milburn:2015}: 
    \begin{align}
        \Gamma = \lim_{\tilde\tau \to \infty} \frac{1}{\tilde\tau} E[|\delta \phi(\tilde \tau)|^2] \label{eq.dephasing_rate_def}
    \end{align} where $E[\cdot]$ denotes ensemble average and $\delta\phi(\tilde \tau)$ refers to the phase when the noise source lasts for a time $\tilde \tau$.
    \item If the statistical properties of the noise considered are given in the frequency space, then transform equation eq.\eqref{eq.general phase fluctuation} to frequency space. Then, the dephasing rate is given by~\cite{Wiener:1930,Khintchine1934,Sneha_HP}: 
    \begin{align}
        \Gamma \leq \int_{-\infty}^{\infty} \mathcal{S}(\omega) \left| \sum_i \sqrt{F_i(\omega)} \right|^2 \, d\omega \label{eq.dephasing rate condensed}
    \end{align}
    where $\mathcal{S}(\omega)$ is the noise power spectral density, which from the Wiener-Khinchin theorem~\cite{Wiener:1930,Khintchine1934} is defined as~\footnote{The notation we use is: $\delta\tilde{\zeta}_{\tilde \tau}(\omega) = \frac{1}{2\pi}\int_{-\tilde \tau}^{\tilde \tau}{\delta\zeta}(t)\ e^{i\omega t}\,dt$}:
    \begin{align}
        \mathcal{S}(\omega)  &=  \lim_{\tilde\tau \to \infty}\frac{1}{\tau}E[\delta\tilde{\zeta}_{\tilde \tau}(\omega)\delta\tilde{\zeta}_{\tilde \tau}^*(\omega)].
    \end{align} 
    \item We compute the effective transfer functions~\footnote{We note that the units of the effective transfer functions is the inverse of that of the power spectral density. The units of the PSD depends depends only on the parameter being perturbed, $\delta \zeta (t)$/ noise source.} $F_i(\omega)$, e.g., $F_1(\omega) \propto \left| \int_{t_i}^{t_f} (x_R - x_L) e^{i \omega t} \, dt \right|^2$, $F_2(\omega) \propto \left| \int_{t_i}^{t_f} (x_R + x_L) e^{i \omega t} \, dt \right|^2$, using the expression obtained {after substituting Eq.\eqref{eq.general phase fluctuation}} in Eq.\eqref{eq.dephasing_rate_def}\cite{Sneha_HP,Sneha_IHP}. {This is illustrated explicitly in the particular case of acceleration and tilt angle fluctuations in the next sections.}
\end{itemize}

The statistical properties of noise are conveniently characterised in frequency space. The first moment gives the mean noise amplitude, the second-order correlation determines the power spectral density, and higher-order correlations encode non-Gaussian features of the noise~\footnote{The Fourier space contains positive and negative frequencies. Mathematically, both are required to describe an arbitrary complex function in the time domain. To obtain a real function in the time domain, the Fourier conjugate must follow: $\tilde\zeta(-\omega) = \tilde\zeta^*(\omega)$. Thus, the negative-frequency components are not independent.}. In the case of Gaussian noise, we have only the first two correlation functions which are independent: 
\begin{align}
    \delta \tilde\zeta(\omega) &= \frac{1}{2\pi}\int_{-\infty}^{\infty}{\delta\zeta}(t)\ e^{i\omega t}\,dt \label{eq.flucFT}\\
    E[\delta\tilde\zeta(\omega)] &= 0 \label{eq.noise_stat1}\\    E[\delta\tilde\zeta(\omega)\delta\tilde\zeta^*(\omega')] &= \mathcal{S}_{\zeta\zeta}(\omega)\delta(\omega-\omega')\label{eq.noise_stat2}
\end{align}
where $E[\cdot]$ denotes ensemble average. $\mathcal{S}_{\zeta\zeta}(\omega)$ is known as the power spectral density of the noise~\cite{Milburn:2015}.

In this paper, we present the dephasing rate due to fluctuations in system parameters, including acceleration ($a$) and tilt angle ($\theta_0$). We  assume white noise statistics, where the power spectral density is a constant:
\begin{align}
    E[\delta\tilde\zeta(\omega)] = 0\,,~~~ %\label{eq.whitenoise_stat1}\\
    \mathcal{S}_{\zeta\zeta}(\omega) \equiv \mathcal{S}_{\zeta\zeta} = \rm const. \label{eq.whitenoise_stat2}
\end{align}

Our aim is to compute the bound on the noise parameter $\mathcal{S}_{\zeta\zeta}$ to obtain a minimum coherence required for spin readout at the end of the interferometer. To obtain quantitative bounds on noise we need to quantify coherence, which is given by:
\begin{equation}
    \text{Coherence measure} = \rm CM = e^{-\Gamma\tau} \in [0,1] \label{eq.cohdef}
\end{equation}
where $\Gamma$ is the dephasing rate and $\tau$ is the time required to complete one-loop interferometer, i.e. one run of the experiment. Note that when the dephasing rate is zero, the coherence measure is unity, and when the dephasing rate tends to arbitrarily large values, the coherence measure vanishes. According to eq.\eqref{eq.cohdef}, the average decoherence time is:
\begin{equation}
    \langle \rm T_{\rm decoh}\rangle = \int_0^\infty CM(\tau)\, d \tau = \int_0^\infty e^{-\Gamma\tau}\, d \tau = {\Gamma}^{-1}
\end{equation} 
One generally imposes that the average decoherence time ${1}/{\Gamma}$ to be greater than the time of the experiment: 
\begin{align}
    {\Gamma}^{-1}>\tau \implies \Gamma \tau <1\implies e^{-\Gamma \tau}>e^{-1}\label{eq.cons_gen}
\end{align}
This constraint gives us a tolerable range of coherence measure as defined in Eq.\eqref{eq.cohdef} to be $[0.37,1]$.

We now compute the dephasing rate introduced by fluctuations in acceleration, and compute the bound on the noise parameters for the specific case of fluctuations in acceleration for the constraint imposed by Eq.\eqref{eq.cons_gen}. We consider various configurations of the acceleration magnitude and direction. %{\bf AM: this subsection is too small, we should think of exapnding it a bit other wise merge them in one section, by removing section A.}

\section{Fluctuations in acceleration}

Consider the fluctuation: $a\to a+\delta a_j(t)$. We consider the fluctuation in the acceleration, $\delta a_j(t)$, to be a function of space, and hence a function of the interferometer arms, $j\in \{R,L\}$. The modified Lagrangian is:
\begin{align}
    \delta L_j = &  m \dot{x}_j\delta\dot{x}_j - m \omega_0^2 x_j \delta x_j - C_j\eta_0\delta x_j \nonumber \\
    &+ ma\delta x_j\cos\theta_0 + m\delta a_jx_j\cos\theta_0\label{eq.lag_fluc_a}
\end{align}
where $\delta x_j$ is the fluctuation introduced in each trajectory due to the fluctuations in the acceleration \footnote{$x_j+\delta x_j$ is the trajectory obtained from the equations of motion derived from the modified Lagrangian, $L_j+\delta L_j$.}. The corresponding fluctuation in the phase, following eq.\eqref{eq.general phase fluctuation} is 
    $\delta \phi = \frac{1}{\hbar}\int_0^\tau (\delta L_R - \delta L_L)\,dt$
where we consider $\tau={2\pi}/{\omega_0}$ indicating a one loop interferometer, where $\omega_0$ is as defined in eq.\eqref{eq.omega0_def}. In Ref.~\cite{Wu:2024tcr}, it was shown that, assuming ideal closure of a harmonic-oscillator-type interferometer, trajectory and velocity fluctuations in Lagrangians of the form of Eq.~\eqref{eq.Lag_j_HP} do not contribute to the dephasing in a one-loop interferometer. This is explicitly shown in Appendix \ref {App.fluc contribution}. Hence, the phase fluctuation arises only from the last term in eq.\eqref{eq.lag_fluc_a}:
\begin{align}
    \delta \phi_a &= \frac{m\cos\theta_0}{\hbar}\int_0^\tau  dt \,\bigg(\delta a_R(t)x_R(t)-\delta a_L(t)x_L(t)\bigg)\label{eq.phase_fluc_a}
\end{align}
Now, we Fourier transform the fluctuations of the acceleration into the frequency basis:
\begin{align}
    \delta a_j(t) &= \int_{-\infty}^{\infty}\,d\omega\,\, \delta \tilde{a}_j(\omega)\,e^{i\omega t}\,.\label{eq.delta_aj_FT}
\end{align}
Further, we will consider an uncorrelated Gaussian noise statistics, as given in eqs.\eqref{eq.noise_stat1}-\eqref{eq.noise_stat2}, in both the arms of the interferometer:
\begin{align}
    E[\delta \tilde{a}_j(\omega)] &= 0 \label{eq.noise_stat_afluc}\\
    E[\delta \tilde{a}^*_j(\omega)\delta \tilde{a}_{j'}(\omega')] &= \mathcal{S}_{aaj}(\omega)\delta(\omega-\omega')\delta_{jj'} \quad j,j'\in\{L,R\} \nonumber
\end{align}
where $\mathcal{S}_{aaj}(\omega)$ is the PSD of the noise considered. By considering $\delta_{jj'}$ term in eq.\eqref{eq.noise_stat_afluc}, we are assuming that the acceleration fluctuations experienced by one arm of the interferometer are uncorrelated to those in the other arm~\footnote{Correlation of noise statistics between the two arms will contribute to dephasing. It is likely that the two arms will have correlated noise statistics since the source of fluctuation is the same. If we assume the noise source is far from the interferometer compared to the distance between the interferometer arms, then the correlation is likely positive, leading to a decrease in the dephasing rate. However, we are not dealing with any particular model for the noise source, and hence compute the dephasing for the uncorrelated case to obtain the upper bound on the dephasing rate given the system parameters. Refer to Appendix.\ref{App.cor_stat}. { For noise analysis of completely correlated acceleration fluctuations, refer \cite{Wu:2024tcr}.}}. In our analysis, we assume that both interferometer arms experience acceleration noise with the same PSD, so that \(\mathcal{S}_{aaR}(\omega)=\mathcal{S}_{aaL}(\omega)\equiv \mathcal{S}_{aa}(\omega)\).

The second order correlation of the phase fluctuation is calculated by imposing the statistics and the equations of motion, i.e. eqs.\eqref{eq.noise_stat_afluc},\eqref{eq.det_traj}, and using eqs.\eqref{eq.phase_fluc_a},\eqref{eq.delta_aj_FT}, to compute the dephasing rate, $\Gamma_{aa}$ upto first order perturbations in the acceleration, see eq.\eqref{eq.dephasing_rate_def}:
\begin{align}
    \Gamma_{aa} &= \lim_{\tilde \tau\to\infty}\frac{1}{\tilde \tau} E[\delta \phi_a^* (\tilde \tau)\delta \phi_a(\tilde \tau)]\nonumber\\
    &= \frac{\cos^2\theta_0}{\omega_0^4} (\tilde{C}_R^2 + \tilde{C}_L^2)\int_{-\infty}^{\infty}d\omega\,\mathcal{S}_{aa}(\omega)\int_0^\tau  dt'\int_0^\tau dt\nonumber\\
    &\quad\times \,e^{i\omega (t-t')} (\cos(\omega_0 t) - 1)(\cos(\omega_0 t') - 1)\nonumber\\
    &= 4\frac{\cos^2\theta_0}{\omega_0^5} (\tilde{C}_R^2 + \tilde{C}_L^2)\times\int_{-\infty}^{\infty}d\xi \,\mathcal{S}_{aa}(\omega_0\xi)\;\frac{\sin^{2}\!(\pi\xi)}{\xi^2(\xi^{2}-1)^{2}}
    \label{eq.second_cor_acc}
\end{align}
where $\xi = {\omega}/{\omega_0}$. By using the expression for the trajectories from eq.\eqref{eq.det_traj}, and comparing the above with the form in eq.\eqref{eq.dephasing rate condensed}, we can obtain the effective transfer function up to first order in acceleration, which is given by:
\begin{align}
    F_{aa,\text{eff}}(\omega)&=  \frac{\cos^2\theta_0}{\omega_0^4} (\tilde{C}_R^2 + \tilde{C}_L^2)\int_0^\tau  dt'\int_0^\tau dt\,\nonumber\\
    &\quad \times(\cos(\omega_0 t) - 1)(\cos(\omega_0 t') - 1)e^{i\omega (t-t')} \nonumber\\
    F_{aa,\text{eff}}(\omega_0\xi) &=\frac{4\cos^2\theta_0}{\omega_0^6} (\tilde{C}_R^2 + \tilde{C}_L^2)\frac{\sin^{2}\!(\pi\xi)}{\xi^2(\xi^{2}-1)^{2}}\,.
    \label{eq.eff_transfun_a}
\end{align} 
To understand the susceptibility of the system to different frequencies of fluctuations in acceleration, we analyse the transfer function given in eq.\eqref{eq.eff_transfun_a}. Note that the coefficient of $f_{aa}(\xi) = {\sin^{2}\!(\pi\xi)}/({\xi^2(\xi^{2}-1)^{2}})$ is a constant for a given set of system parameters like $\theta_0$, $a$, $\omega_0$, etc. Hence, to understand the bandwidth of frequencies that can affect the system, it is sufficient to look at the behaviour of $f_{aa}(\xi)$.
%%%%%%%%%%%%%%%%%%%%%%%%%
\begin{figure}[ht!]
    \centering
    \includegraphics[width=\linewidth]{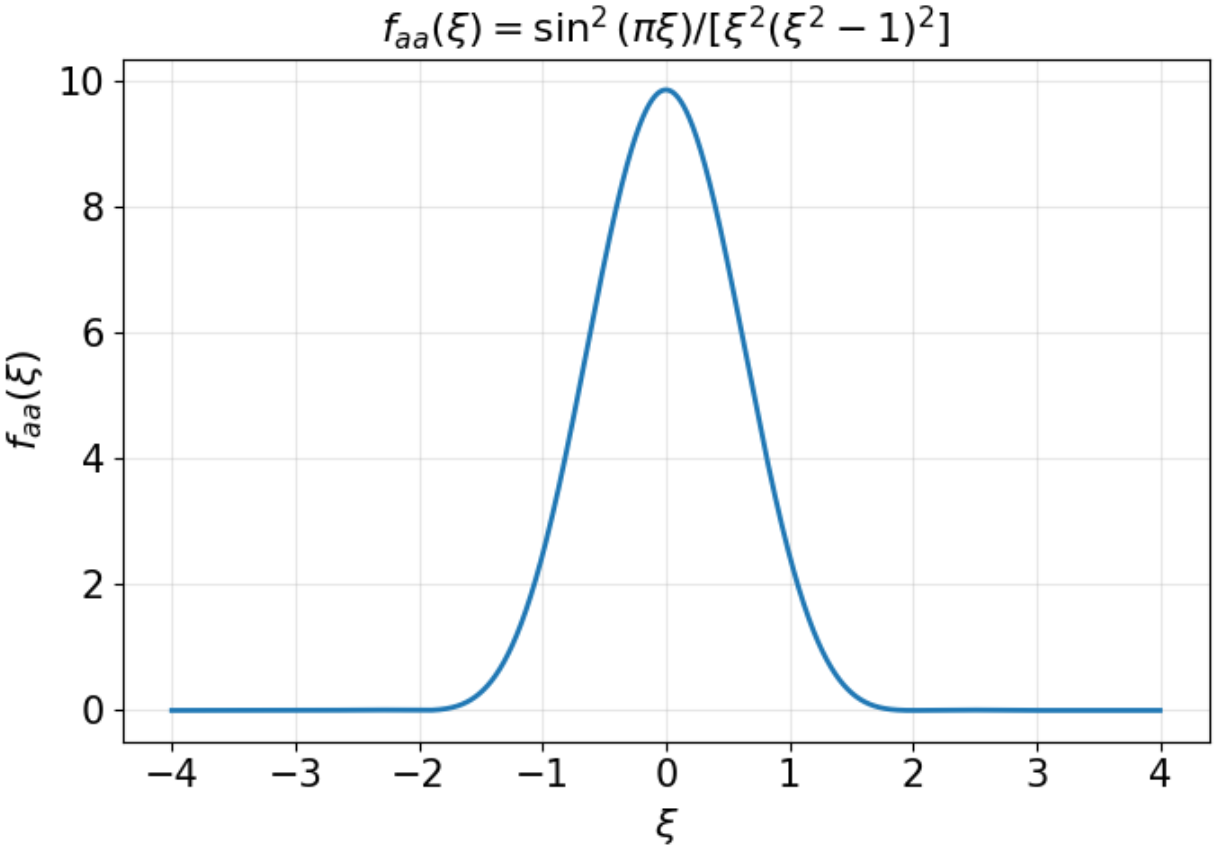}
    \caption{The plot shows the frequency response of the interferometer to acceleration fluctuations. The horizontal axis is the dimensionless quantity $\xi = \omega/\omega_0$, where $\omega$ is the angular frequency of the noise source and $\omega_0$ is the characteristic angular frequency of the harmonic oscillator. The vertical axis shows $f_{aa}(\xi)$, the frequency-dependent weighting factor appearing in the acceleration-noise transfer function eq.\eqref{eq.eff_transfun_a}. Larger $f_{aa}(\xi)$ indicates greater sensitivity to perturbations at that normalised frequency.}
    \label{fig:f_aa(xi)}
\end{figure}
%%%%%%%%%%%%%%%%%%%%%%%%%%%
From Fig.\ref{fig:f_aa(xi)}, we can understand that when the noise source is white, i.e., where the PSD is constant over all frequencies, the dephasing rate involving the product of the PSD and the transfer function makes it integrable over the whole range of frequencies. We further note that the values of $\xi$ contributing significantly to the transfer function is $|\xi|\leq1\implies|\omega|\leq|\omega_0|$.

In many experiments, one can argue that when the experimental time duration is $\tau={2\pi}/{\omega_0}$, noise components with frequencies much less than $\omega< \omega_0$ can be neglected, since they are effectively constant over this time interval. This reasoning is only valid for observables that are insensitive to quasi-static offsets, i.e., the system's transfer function is negligibly small at low frequencies. However, from the transfer function obtained (see Fig.\ref{fig:f_aa(xi)}), one can see that this is not true. The system is susceptible to non-negligible perturbations even from small frequency noise.

\section{White noise analysis for the acceleration}

Now we consider white noise for the external acceleration, i.e.
\begin{align}
    \mathcal{S}_{aa}(\omega) \equiv \mathcal{S}_{aa} = \rm constant
\end{align}
Thus, the dephasing rate, using eq.\eqref{eq.second_cor_acc} becomes:
\begin{align}
    \Gamma_{aa} &= 4\frac{\cos^2\theta_0}{\omega_0^5} (\tilde{C}_R^2 + \tilde{C}_L^2)\times\frac{3\pi^2}{2}\mathcal{S}_{aa}\,.\label{eq.deph_rate_acc}
%     &= 4\frac{\sin^2\theta_0}{\omega_0^5} (\tilde{C}_R^2 + \tilde{C}_L^2)\times\frac{3\pi^2}{2}\mathcal{S}_{aa}
\end{align}
%
\begin{comment}
    We consider the experimental setup described by the parameter values described in Table.\ref{}.
\begin{table}[ht!]
\centering
\begin{tabular}{|W{c}{0.2\linewidth}|W{c}{0.4\linewidth}|}
\hline
Parameter & Value \\
\hline
$\eta_0$ & $6\times 10^3~\mathrm{T\,m^{-1}}$ 
\\
$m$ & $10^{-15}~\mathrm{kg}$ 
\\
$T$ & $2\pi/\omega_0\simeq 0.015~\mathrm{s}$ 
\\
\hline
\end{tabular}
\caption{Parameter values used to evaluate acceleration-noise bounds.}\label{tab:params_acc_noise}
\end{table}
\end{comment}
For purpose of illustration, we consider a particular experimental setup with the magnetic field gradient, which determines the characteristic frequency, $\omega_0$, and $\tau$, see eq.\eqref{eq.omega0_def}, to be $\eta_0 = 6\times 10^3\,\text{Tm}^{-1}$, similar to values assumed in \cite{Marshman:2021wyk,Sneha_HP,Sneha_IHP}. For a one-loop interferometer, the one-run experimental time is $\tau = {2\pi}/{\omega_0}\simeq0.015$s for a mass $m=10^{-15}{\rm kg}$ yields 
$\Delta x=1$nm. 

We further consider three scenarios: (a) The random acceleration fluctuation is along the interferometer axis where the magnitude of mean acceleration varies, (b) the external acceleration is assumed here to be zero ($a=0~{\rm m s^{-2}}$). Further, we also consider the random acceleration fluctuations is at an arbitrary angle with respect to the interferometer axis, and (c) the mean acceleration is the gravitational acceleration, and $\theta_0\sim 90^\circ$\footnote{For several fluctuation sources $\{\delta\zeta_j\}_{j=1}^n$, with individual dephasing rates $\Gamma_j$, the total dephasing can be conservatively bounded without assuming decorrelation by applying the triangle inequality to the corresponding phase fluctuations:
\[
\Gamma \leq \left(\sum_{j=1}^n \sqrt{\Gamma_j}\right)^2 .
\]
}. %Note that the upper bound on the total dephasing rate due to acceleration fluctuations from different sources can be computed using eq.\eqref{eq.multisource_fluc}.

\subsection{Acceleration fluctuations along the interferometer axis}
\begin{figure}[ht!]
    \centering
    \includegraphics[width=\linewidth]{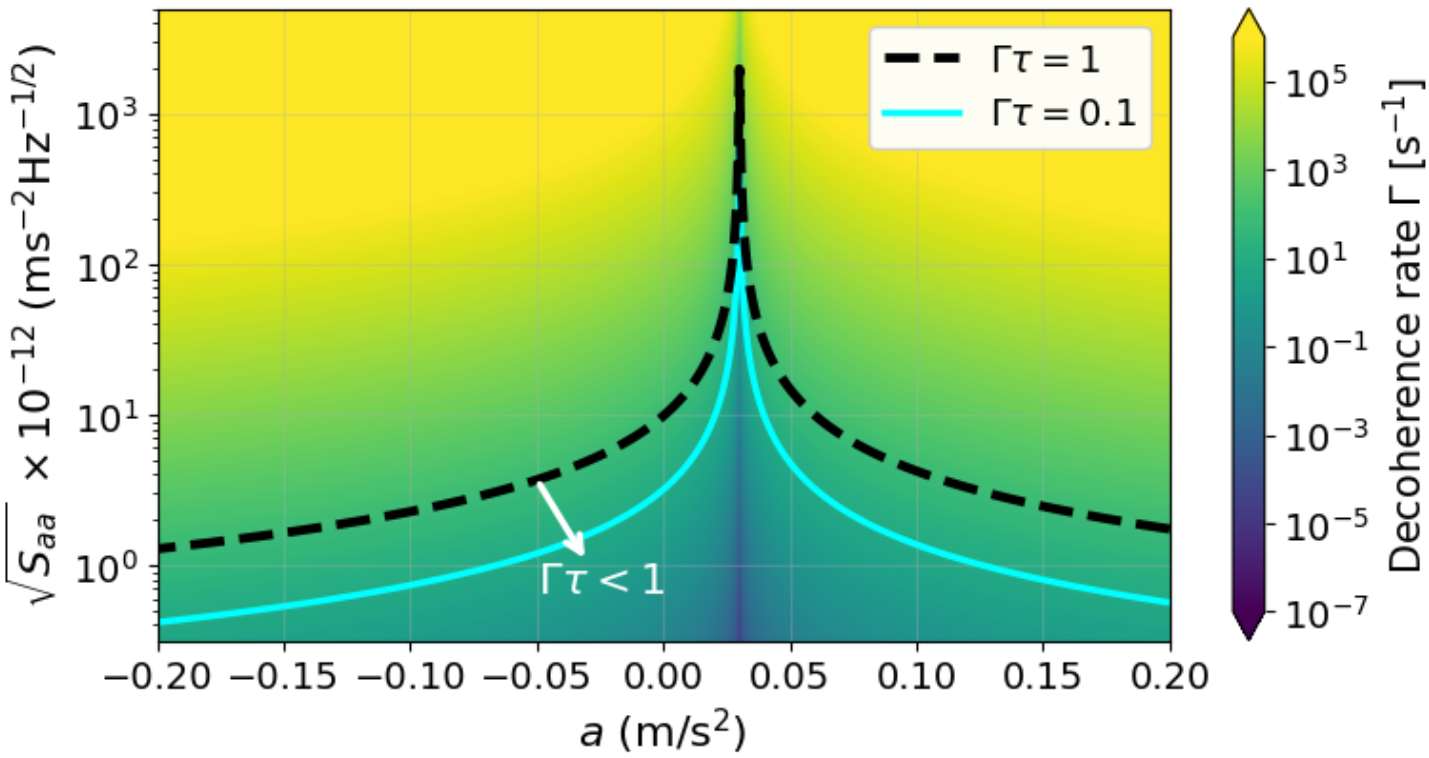}
    \caption{The contour plots map the dephasing rate \(\Gamma\) as a function of the white noise amplitude spectral density \(\sqrt{\mathcal{S}_{aa}}\) and the external acceleration magnitude $a$ with all other parameters fixed to the values: $\theta_0 = 0^\circ, ~\eta_0=6\times 10^{3}{\rm Tm^{-1}}, m=10^{-15}{\rm kg},~ \tau=2\pi/\omega_0\simeq0.015$s, which yields $\Delta x=1$nm. The region below the black dashed contour corresponds to the experimentally desirable operating regime satisfying \(\Gamma \tau < 1\). The graph shows that the behaviour of the dephasing rate and the tolerable square root of PSD, $\sqrt{\mathcal{S}_{aa}}$ is finite and smooth at $a_m\sim 0.03 \ \rm m s^{-2}$.}
    \label{fig:rtS_aa_vs_a_-02-02}
\end{figure}

In Fig.\ref{fig:rtS_aa_vs_a_-02-02}, we observe that the tolerable noise amplitude increases with the decrease in magnitude of acceleration, $a$. This corroborates with eq.\ref{eq.deph_rate_acc}, in the regime when the external force $ma\cos\theta_0>C_j\eta_0$. However, 
when $ma\cos\theta_0\approx C_j\eta_0$, the square root of the PSD acquires a maximum value for a fixed dephasing rate. Note that this value may change if we were to take higher order perturbations in the acceleration.
%the true minima in tolerance of the system to acceleration noise is achieved for the value of acceleration such that $\tilde{C}_R^2+\tilde{C}_L^2$ is minimized for a fixed $\theta_0$. 
For $\theta_0 = 0^\circ$, this is achieved for $a = a_m \sim 0.03\ \rm ms^{-2}$. The smoothness of the plot in this regime is shown in Appendix.\ref{App.zoom}. In particular, for $\Gamma \tau<1$, we have the following bound on the noise amplitude spectral density:$$
    \sqrt{\mathcal{S}_{aa}} < {\cal O}({10^{-8}}){\rm m s^{-2} Hz^{-1/2}}, {\rm for}~~ a = a_m =0.03\ \rm ms^{-2}.$$

For illustration only, we also present the bound on the acceleration noise amplitude under the constraint $\Gamma \tau<1$, when the acceleration magnitude is $a=3~{\rm ms^{-2}}$ and $\theta_0 = 0^\circ$: $$
    \sqrt{\mathcal{S}_{aa}} < {\cal O}({10^{-13}}){\rm m s^{-2} Hz^{-1/2}}, \ {\rm for} \ a = 3{\rm ms^{-2}}$$.

\subsection{Arbitrary acceleration fluctuations with zero mean}

\begin{figure}[ht!]
    \centering
    \includegraphics[width=\linewidth]{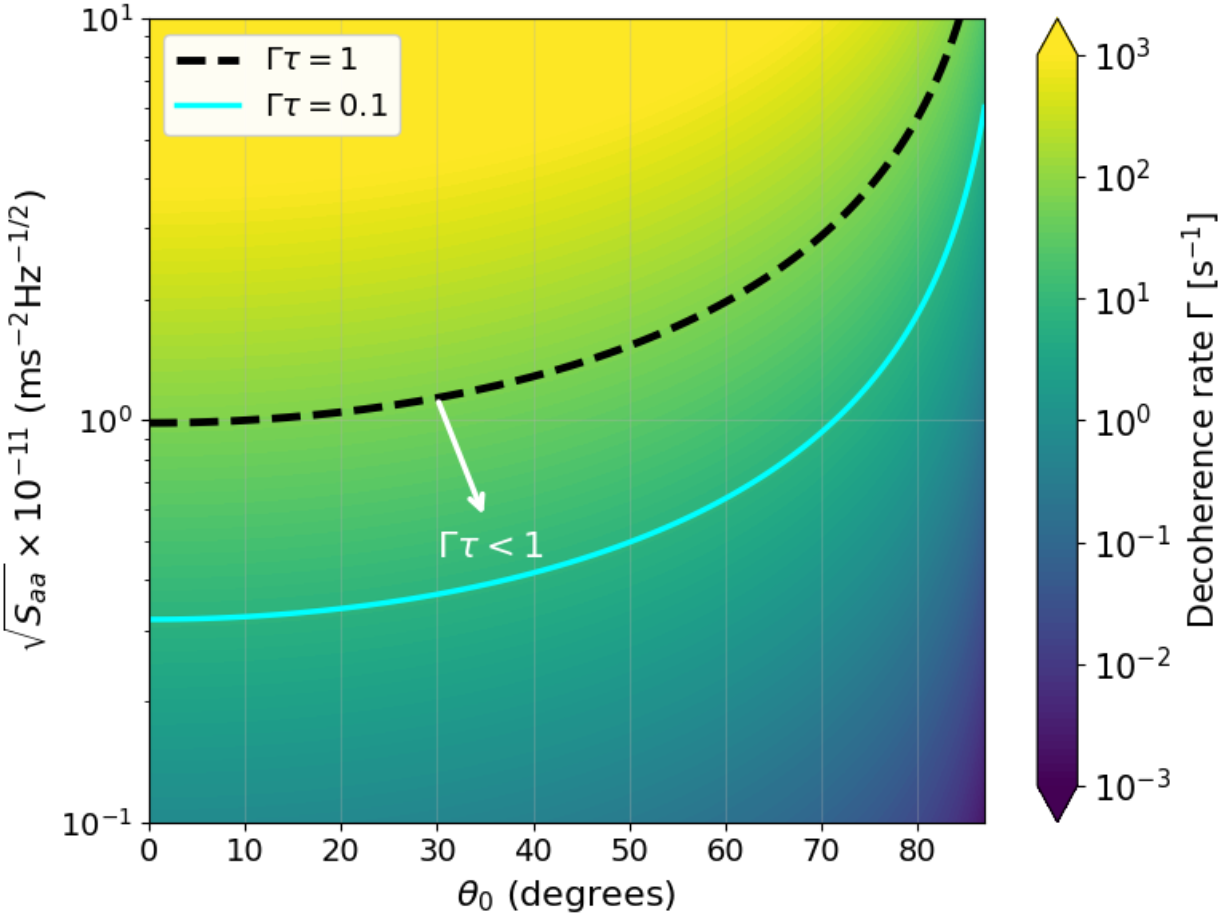}
    \caption{The contour plot maps the dephasing rate \(\Gamma\) as a function of the white noise amplitude spectral density \(\sqrt{\mathcal{S}_{aa}}\) and the tilt angle \(\theta_0\) with all other parameters fixed to the values: (no external acceleration) $a = 0 {\rm ms^{-2}}$,~$\eta_0=6\times 10^{3}{\rm Tm^{-1}}$, $m=10^{-15}{\rm kg}$, $\Delta x=1$nm, $\tau=2\pi/\omega_0\simeq0.015$s. The region below the black dashed contour corresponds to the experimentally desirable operating regime satisfying \(\Gamma \tau < 1\).}
    \label{fig:rtS_aa_vs_theta_0_0-87}
\end{figure}

We consider the setup when the mean acceleration is zero ($a=0$) and the acceleration fluctuation both are along the interferometer axis. In Fig.\ref{fig:rtS_aa_vs_theta_0_0-87}, we observe that the tolerable noise amplitude decreases with the decrease in tilt angle, $\theta_0$. This corroborates with eq.\ref{eq.deph_rate_acc}. This aligns with the physical intuition that when the acceleration fluctuations align along the interferometer axis, the dephasing is higher. While, if the acceleration fluctuations are perpendicular to the interferometer axis, the dynamics along the two axis decouple and hence ideally, no dephasing is expected. Hence, in the limit $\theta_0\to90^\circ$, we observe that an arbitrary noise amplitude is tolerable. However, this will not be the case, if we were to consider any dynamics of the nanodiamond in the direction of gravity.

In particular, for $\Gamma \tau <1$, we have the following bound on the noise amplitude spectral density when the fluctuations are along the interferometer axis ($\theta_0=0^\circ$):
$$\Gamma \tau< 1 \implies \sqrt{\mathcal{S}_{aa}} < \order{10^{-11}}\rm m s^{-2} Hz^{-1/2}.$$

%%%%%%%%%%%%%%%%%%%%%%%%%%%%%%%%%

\begin{comment}
    
The power spectral density of the noise component along the interferometer axis will be $\mathcal{S}_{aa}\cos^2{\theta_0}$. In Fig.\ref{fig:rtS_aac_vs_ac_-3-3}, we plot the dephasing rate with respect to variation of the amplitude spectral density along the interferometer axis, $\sqrt{\mathcal{S}_{aa}}\ \cos{\theta_0}$ and the acceleration component along the interferometer axis.

\begin{figure}[ht!]
    \centering
    \includegraphics[width=\linewidth]{rtS_aac_vs_ac_-3-3.png}
    \caption{Caption}
    \label{fig:rtS_aac_vs_ac_-3-3}
\end{figure}

\begin{figure}[ht!]
    \centering
    \includegraphics[width=\linewidth]{rtS_aac_vs_ac_zoom.png}
    \caption{Caption}
    \label{fig:placeholder}
\end{figure}

\begin{figure}[ht!]
    \centering
    \includegraphics[width=\linewidth]{rtS_aa_vs_theta_0_0-80.png}
    \caption{a=0}
    \label{fig:placeholder}
\end{figure}

\end{comment}

\subsection{Gravitational fluctuations - Acceleration nearly perpendicular to interferometer axis}
In general, we create the spatial superposition along an axis perpendicular to gravity. However, the interferometer axis may have a small tilt along the gravitational axis. In this subsection, we analyse the tolerable acceleration fluctuations along the gravitational axis. 

\begin{figure}[ht!]
    \centering
    \includegraphics[width=\linewidth]{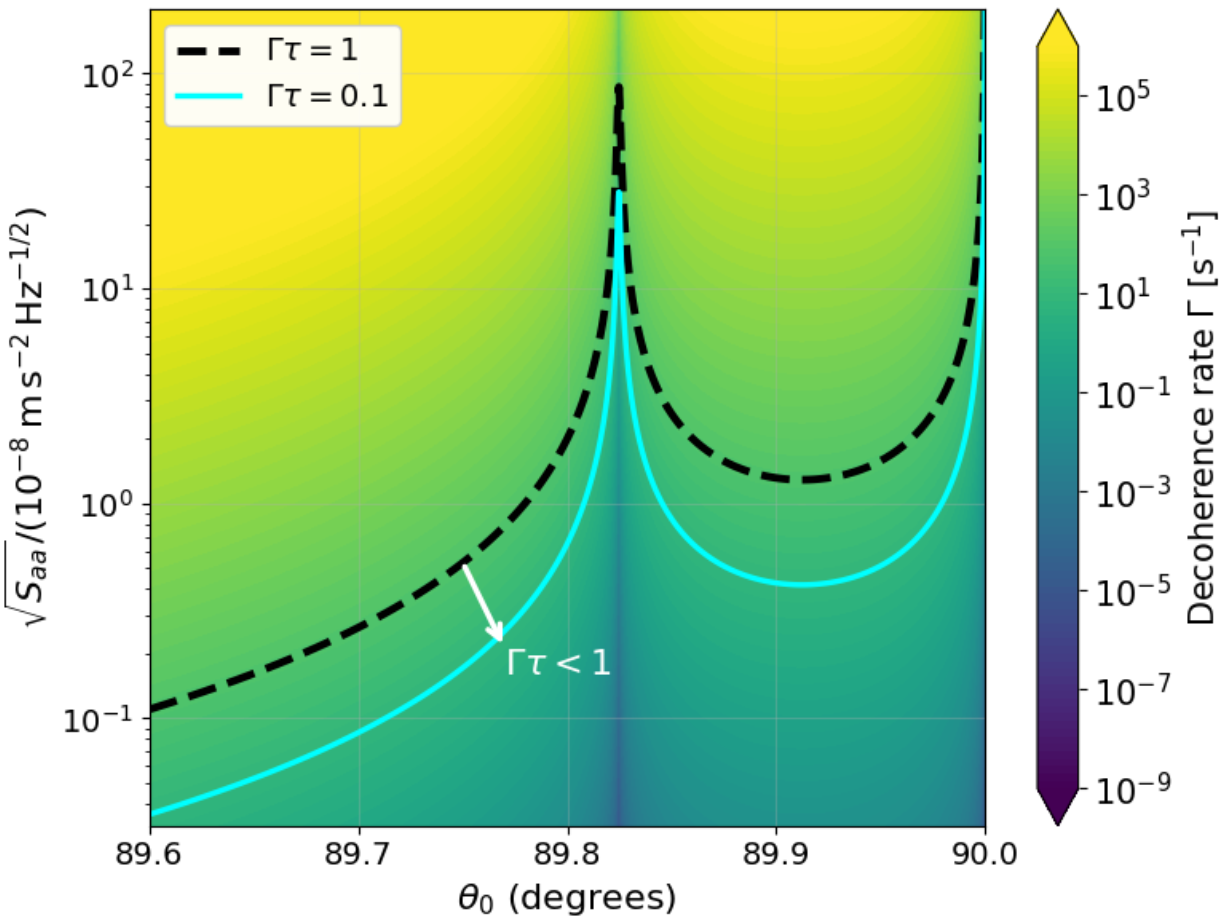}
    \caption{The contour plots map the dephasing rate \(\Gamma\) as a function of the white noise amplitude spectral density \(\sqrt{\mathcal{S}_{aa}}\) and the tilt angle \(\theta_0\) with all other parameters fixed to the values: $a = 9.81 {\rm ms^{-2}},~\eta_0=6\times 10^{3}{\rm Tm^{-1}}$, $m=10^{-15}{\rm kg}$, $\Delta x=1$nm, $\tau=2\pi/\omega_0\simeq0.015$s. The region below the black dashed contour corresponds to the experimentally desirable operating regime satisfying \(\Gamma \tau < 1\). When the acceleration fluctuations are perpendicular to the interferometer axis ($\theta_0=90^\circ$), the dynamics along the two axes decouple and hence an arbitrary amplitude of $\sqrt{\mathcal{S}_{aa}}$ is tolerable at exactly $\theta_0=90^\circ$. The tolerable noise amplitude spectral density, $\sqrt{\mathcal{S}_{aa}}$ is finite and smooth at $\theta_0\sim 89.825^\circ$, where a finite local minima in tolerance of the system to acceleration noise is achieved.}
    \label{fig:rtS_aa_vs_theta_9.8_89.6-90}
\end{figure}

In Fig.\ref{fig:rtS_aa_vs_theta_9.8_89.6-90}, we observe that there is no constraint on the amplitude power spectral density when the acceleration fluctuations are perpendicular to the interferometer axis. This is because the dynamics and the fluctuations in the acceleration along the two axis decouple. However, for within a degree of deviation from the perpendicular orientation, we observe the following bound on $\sqrt{\mathcal{S}_{aa}}$ corresponding to the constraint $\Gamma < \tau^{-1}$, and $a=g=9.81~ {\rm ms^{-2}}$:
\begin{align}
     \sqrt{\mathcal{S}_{aa}} &< \order{10^{-10}}\rm m s^{-2} Hz^{-1/2} \quad \rm{for}\quad \theta_0 = 89^\circ\nonumber \\
     \sqrt{\mathcal{S}_{aa}} &< \order{10^{-6}}\rm m s^{-2} Hz^{-1/2} \quad \rm{for}\quad \theta_0 \sim 89.825^\circ \nonumber
\end{align}
We observe that the bound on the $\sqrt{\mathcal{S}_{aa}}$ changes by four orders of magnitude for a change within $1^\circ$ of tilt angle. Thus, to operate in a regime where the system is less sensitive to the acceleration fluctuations, we need to fix the tilt angle between $89.8^\circ$ and $90^\circ$. The smoothness of the plot in this regime is shown in Appendix.\ref{App.zoom}.

In Fig.\ref{fig:rtS_aa_vs_theta_9.8_89.6-90}, to the right of $\theta_0\sim89.825^\circ$, the magnetic force, $-({\chi_\rho m}/{\mu_0})B_{0}\eta_0$ dominates over external acceleration $ma\cos\theta_0$. Thus, between $89.9^\circ$ and $90^\circ$, we can Taylor expand the dephasing rate expression given by eq.\eqref{eq.deph_rate_acc} with respect to theta {and hence ignoring the projection of external acceleration term on the interferometer axis, in particular setting $\tilde{C}_j = C_j$}, to obtain:
\begin{align}
    \Gamma_{aa} &= \frac{4\theta_0^2}{\omega_0^5} (C_R^2 + C_L^2)\times\frac{3\pi^2}{2}\mathcal{S}_{aa}\,.\label{eq.deph_rate_acc_nearperp}
\end{align}
{In this regime, when the external acceleration is almost perpendicular to the interferometer axis, the dephasing rate becomes independent of the magnitude of the external acceleration, $a$. 

We now note that} the sensitivity of the system to acceleration noise achieves a local minimum for $\theta_0\sim 89.825^\circ$. This corresponds to $\theta_0$ that minimizes $\tilde{C}_R^2+\tilde{C}_L^2$ ~\footnote{Recall that: $C_j = \bigg( S_{xj}\hbar\gamma_e - \frac{ \chi_\rho m}{\mu_0}B_{0}\bigg)$ and $\tilde{C}_j = C_j \eta_0-ma\cos\theta_0$. We note that in the term $C_j$, $-({ \chi_\rho m}/{\mu_0})B_{0}$ term is order of two times greater than $S_{xj}\hbar\gamma_e$. Further, the former is interferometer arm independent, while the latter is interferometer arm dependent. Hence, minimizing $\tilde C_j$ in eq.\eqref{eq.tilde_C_j}, corresponds to tuning $ma\cos\theta_0$ to cancel the contribution of $-({ \chi_\rho m}/{\mu_0})B_{0}$. Physically, this occurs because the spin-independent magnetic force ($-({\chi_\rho m}/{\mu_0})B_0\eta_0$) can be tuned to cancel the effective inertial force ($ma\sin\theta_0$), yielding the net force along the interferometer axis to arise only from spin dependent force $S_{xj}\hbar\gamma_e\eta_0$.}.  In particular, this occurs when:
\begin{align}
    \tilde C_j = S_{xj}\hbar\gamma_e \quad \text{when} \;- \frac{ \chi_\rho m}{\mu_0}B_{0}\eta_0 = ma\sin\theta_0
\end{align}

Conversely, if the noise amplitude is $\sqrt{\mathcal{S}_{aa}} \sim \order{10^{-8}}\rm m s^{-2} Hz^{-1/2}$, the favorable operating regime is with $\theta_0\in(89.8^\circ, 90^\circ)$.

\section{Fluctuations in the tilt of the interferometer}
Consider the fluctuation: $\theta_0 \to \theta_0+\delta \theta(t)$. The procedure is similar to the previous section. The modified Lagrangian is:
\begin{align}
    \delta L_j &= mx_ja_j\sin\theta_0\delta\theta + ...
\end{align}
where $...$ are due to trajectory and velocity fluctuations that do not contribute to dephasing in a one loop interferometer (refer Appendix.\ref{App.fluc contribution}). The corresponding fluctuation in the phase is:
\begin{align}
    \delta \phi_\theta &= \frac{1}{\hbar}\int_0^\tau (\delta L_R - \delta L_L)\,dt\\
    &= \frac{ma}{\hbar}\sin\theta_0\int_0^\tau \delta \theta (x_R - x_L)\,dt
\end{align}
where we consider $\tau=\frac{2\pi}{\omega_0}$ indicating a one loop interferometer. Similar to the previous section, we go to the Fourier basis:
\begin{align}
    \delta \theta(t) &= \int_{-\infty}^{\infty}\,d\omega\,\, \delta \tilde{\theta}(\omega)\,e^{i\omega t}\label{eq.delta_theta_FT}
\end{align}

Now we consider the fluctuations in $\theta_0$ to follow gaussian noise statistics as given in eqs.\eqref{eq.noise_stat1}-\eqref{eq.noise_stat2}.
\begin{align}
    E[\delta \tilde{\theta}_j(\omega)] &= 0\\
    E[\delta \tilde{\theta}^*(\omega)\delta \tilde{\theta}(\omega')] &= \mathcal{S}_{\theta\theta}(\omega)\delta(\omega-\omega')\label{eq.noise_stat2_thetafluc}
\end{align}
where $\mathcal{S}_{\theta\theta}(\omega)$ is the PSD of the noise considered.

The dephasing rate is calculated using its relation to the second order correlation of the phase fluctuation:
\begin{align}
    \Gamma_{\theta\theta} &= \lim_{\tilde \tau\to\infty}\frac{1}{\tilde \tau} E[\delta \phi_\theta^* (\tilde \tau)\delta \phi_\theta(\tilde \tau)]\nonumber\\
    &=\bigg(\frac{a\sin\theta_0}{\omega_0^2}\bigg)^2(2\gamma_e\eta_0)^2\int_0^\tau \int_0^\tau \int_{-\infty}^{\infty} S_{\theta\theta}(\omega)\nonumber\\
    &\quad\times(\cos(\omega_0 t) - 1)(\cos(\omega_0 t') - 1) e^{i\omega (t -t')}\,dt'dt\, d\omega \label{eq.deph_rate_thetafluc}
\end{align}
where $\Gamma_{\theta\theta}$ is the dephasing rate due to fluctuations in the tilt angle, $\theta_0$. 
\begin{comment}
    Under the small $\theta_0$ approximation, the dephasing rate becomes:
\begin{align}
    \Gamma_{\theta\theta} 
    &=\bigg(\frac{a\theta_0^2}{\omega_0^2}\bigg)^2(2\gamma_e\eta_0)^2\int_0^\tau \int_0^\tau \int_{-\infty}^{\infty} S_{\theta\theta}(\omega)\nonumber\\
    &\quad\times(\cos(\omega_0 t) - 1)(\cos(\omega_0 t') - 1) e^{i\omega (t -t')}\,dt'dt\, d\omega \label{eq.small_theta_tilt}
\end{align}
\end{comment}

For small tilt angles, Eq. \eqref{eq.deph_rate_thetafluc} suggests that the dephasing rate becomes negligible. However, this apparent suppression should be interpreted with caution: in this regime the approximation underlying the expression breaks down, because when the leading (first-order) dephasing contribution is small, higher-order effects fluctuations—can become non-negligible and may dominate the dephasing. However, in the analysis, while computing $\delta \phi^* \delta \phi$ in eq.\eqref{eq.deph_rate_thetafluc}, only the terms in the first order of fluctuations were retained.

Upon comparing the eq.\eqref{eq.deph_rate_thetafluc} with the form in eq.\eqref{eq.dephasing rate condensed}, the effective transfer function is given by eq.\eqref{eq.eff_transfun_thetafluc}.
\begin{align}
    |F_{\theta, \text{eff}}(\omega)|^2 &= \bigg(\frac{a\sin\theta_0}{\omega_0^2}\bigg)^2(2\gamma_e\eta_0)^2\int_0^\tau  dt'\int_0^\tau dt\,\nonumber\\
    &\quad \times(\cos(\omega_0 t) - 1)(\cos(\omega_0 t') - 1)e^{i\omega (t-t')} \nonumber\\
    |F_{\theta,\text{eff}}(\omega_0\xi)|^2 &=(4\gamma_e\eta_0)^2\bigg(\frac{a\sin\theta_0}{\omega_0^3}\bigg)^2\frac{\sin^{2}\!(\pi\xi)}{\xi^2(\xi^{2}-1)^{2}} \label{eq.eff_transfun_thetafluc}
\end{align}
We note that the behaviour of the part of the transfer function dependent on the noise frequency $\omega$ is the same as in eq.\eqref{eq.eff_transfun_a}. Hence, the behaviour of the transfer function with respect to $\xi = {\omega}/{\omega_0}$ is explained by Fig.\ref{fig:f_aa(xi)}. Only the scaling factors vary between the transfer functions corresponding to acceleration fluctuations and those corresponding to tilt fluctuations.

\section{White noise analysis for the tilt}

We consider the case of white noise statistics with PSD given by:
\begin{align}
    \mathcal{S}_{\theta\theta}(\omega) = \mathcal{S}_{\theta\theta}.
\end{align}
Similar to the previous section, we consider a particular experimental setup with the magnetic field gradient, which determines the characteristic frequency, $\omega_0$, and $T$, see eq.\eqref{eq.omega0_def}, to be $\eta_0 = 6\times 10^3\,\text{Tm}^{-1}$, similar to values assumed in \cite{Marshman:2021wyk,Sneha_HP,Sneha_IHP}. For a one-loop interferometer, the one-run experimental time is $\tau = {2\pi}/{\omega_0}\simeq0.015$s for a mass $m=10^{-15}{\rm kg}$.

In particular, we now consider the scenario where gravitational acceleration is perpendicular to the axis of spatial superposition ($\theta_0=90^\circ$) in the presence of tilt angle fluctuations with PSD $\mathcal{S}_{\theta\theta}$. The upper bound on $\sqrt{\mathcal{S}_{\theta\theta}}$ under the constraint that $\Gamma_{\theta\theta} \leq {\omega_0}/{2\pi}$ is given by: 
\begin{align}
    \sqrt{\mathcal{S}_{\theta\theta}} \lesssim \order{10^{-10}}\, \text{rad}\,\text{Hz}^{-1/2}\nonumber
\end{align}
Now we shall probe the general trend of the upper bound on $\sqrt{\mathcal{S}_{\theta\theta}}$ for different values of random acceleration, $a$ and for different values of tilt angle, $\theta_0$. We shall also probe the effect of varying the constraint on the coherence.

%%%%%%%%%%%%%%%%%%%%%%%%%%%%%%%%%%%%%%%%%%%%
\begin{figure}[ht!]
    \centering
    \begin{subfigure}[t!]{\linewidth}
        \includegraphics[width=\linewidth]{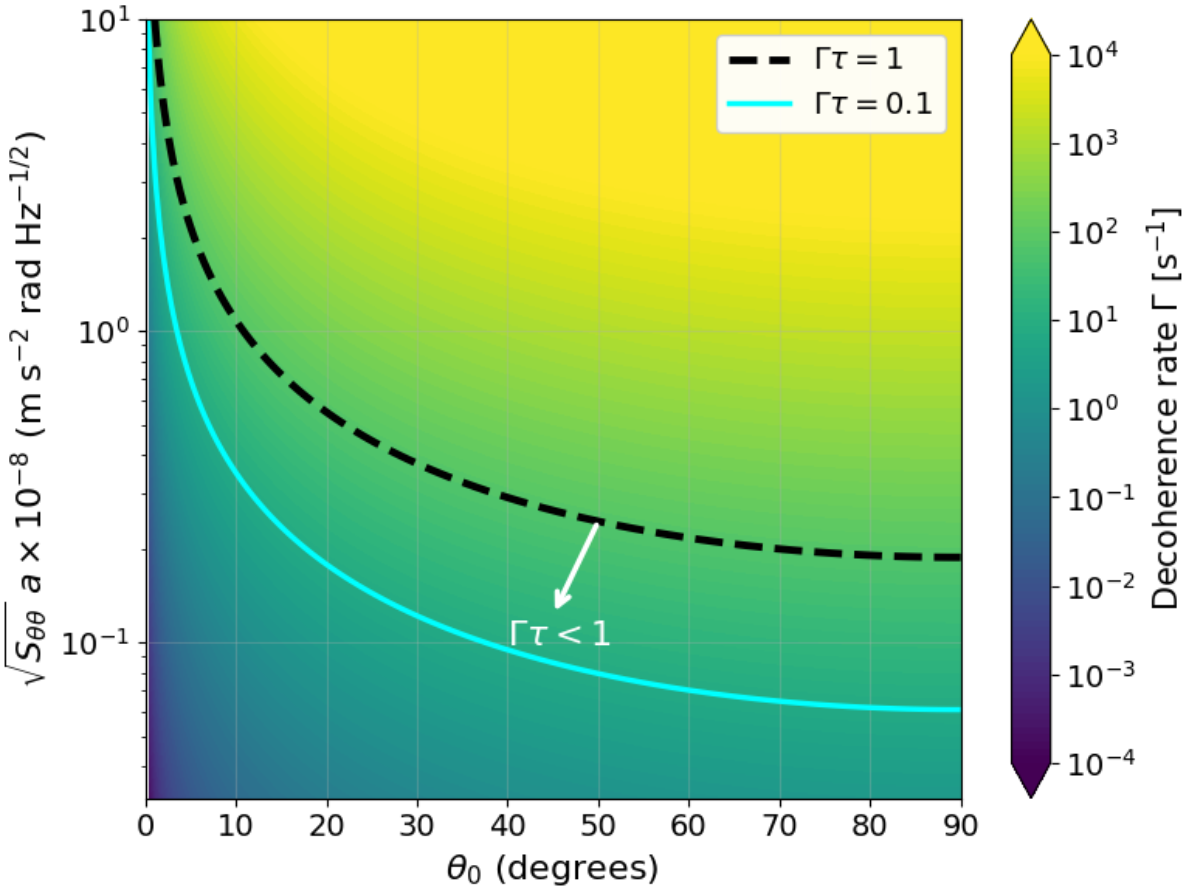}
        \caption{For a fixed dephasing rate and acceleration magnitude, the upper bound on tolerable noise amplitude $\sqrt{\mathcal{S}_{\theta\theta}}$ loosens as the interferometer axis aligns with the direction of the applied acceleration (i.e., as $\theta_0\to 0^\circ$). 
        Note that the results close to $0^\circ$ is not reliable since here, the second order effect of tilt fluctuations on the equation of motion become significant, however, this has not been considered in the present analysis.}
        \label{fig:rtS_tt_vs_theta_arb_0-90}
    \end{subfigure}
    \hfill
    \begin{subfigure}[t!]{\linewidth}
        \includegraphics[width=\linewidth]{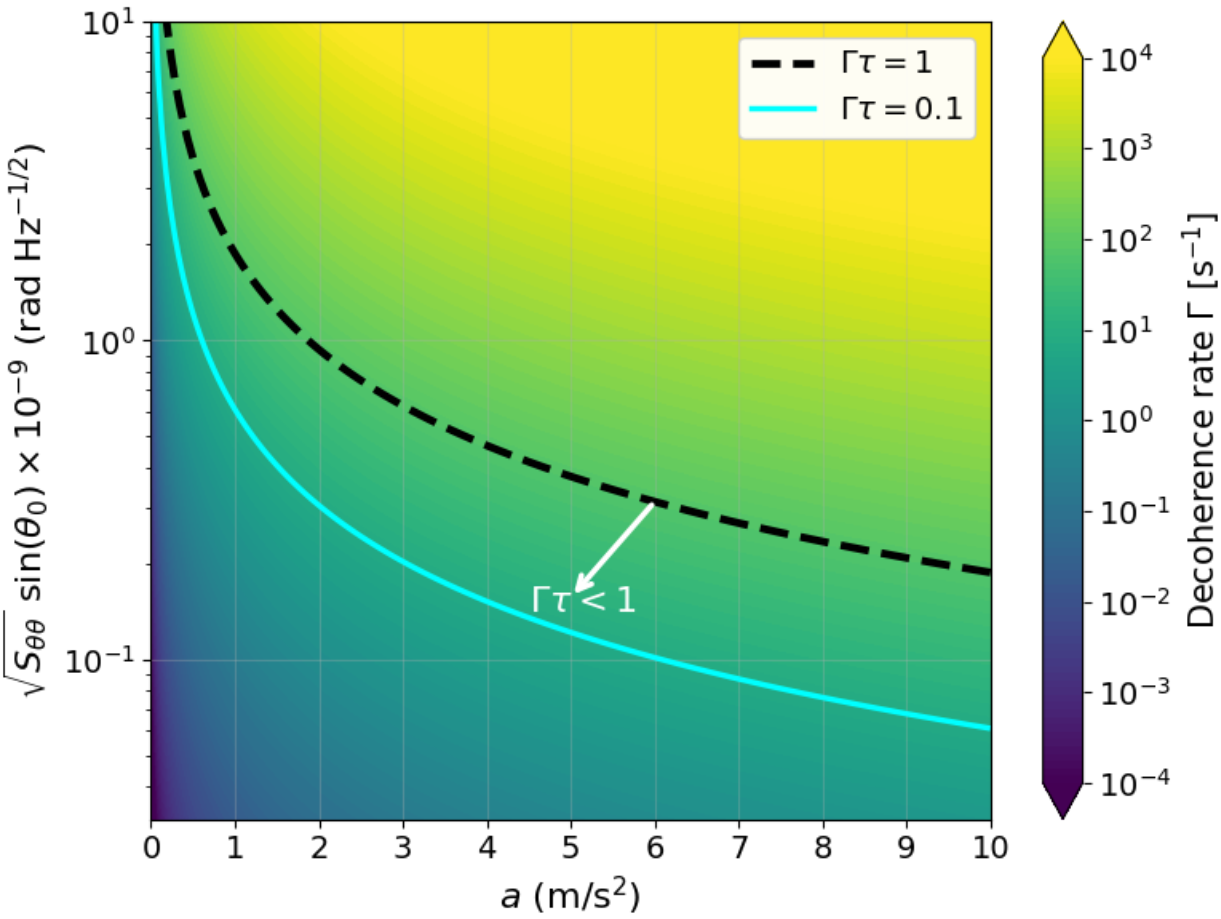}
        \caption{For a fixed dephasing rate, the upper bound on $\sqrt{\mathcal{S}_{\theta\theta}}$ becomes more stringent as the acceleration magnitude increases. In particular, for zero external acceleration, fluctuations in the tilt angle doesn't lead to dephasing.}
        \label{fig:rtS_tt_vs_a_arb_0-10}
    \end{subfigure}
    \caption{The contour plots map the dephasing rate \(\Gamma\) as a function of (a) \(\sqrt{\mathcal{S}_{\theta\theta}}\ a\) and the tilt angle \(\theta_0\), (b) \(\sqrt{\mathcal{S}_{\theta\theta}}\ \sin(\theta_0)\) and acceleration $a$, with all other parameters fixed to the values: $~\eta_0=6\times 10^{3}{\rm Tm^{-1}}$, $m=10^{-15}{\rm kg}$, $\Delta x=1$nm,~$\tau=2\pi/\omega_0\simeq0.015$s. The region below the black dashed contour corresponds to the regime satisfying \(\Gamma \tau < 1\).}
\end{figure}

%%%%%%%%%%%%%%%%%%%%%%%%%%%%%%%%%%%%%%%

In Fig.\ref{fig:rtS_tt_vs_theta_arb_0-90}, we observe that for a fixed dephasing rate and acceleration magnitude, with increase in the tilt angle, the constraint on $\sqrt{\mathcal{S}_{aa}}$ decreases. This corroborates with eq.\eqref{eq.deph_rate_thetafluc}, where the dephasing rate becomes maximum for $\theta_0 = 90^\circ$. Further, the order of magnitude of the bounds for different coherence constraints~\footnote{A coherence measure of $\rm CM = 0.9$ corresponds to $\Gamma \tau\sim 0.1$.}, keeping all other parameters fixed, varies by at most one order of magnitude.

In Fig.\ref{fig:rtS_tt_vs_a_arb_0-10}, we see that the bound on  $\sqrt{\mathcal{S}_{\theta\theta}}$ for a fixed dephasing rate and tilt angle, becomes stricter with increase in acceleration magnitude. Note that the acceleration dependence is introduced in the dephasing rate through the overall factor in eq.\eqref{eq.deph_rate_thetafluc}.

%%%%%%%%%%%%%%%%%%%%%%%%%%%%%%%%%%%%%%%
\section{Conclusion}

 We have presented a quantitative noise analysis for fluctuations in random acceleration and in the tilt angle between the spatial superposition axis and the plane perpendicular to the acceleration in a matter-wave interferometer. We identified an operating regime in which the \emph{spin-independent} contributions to the net force on each arm of the interferometer cancel. In this regime, the system is least susceptible to dephasing due to acceleration noise acting along the direction of the creation of the spatial superposition.

We showed how to quantitatively assess the fluctuations in (i) the random external acceleration $a$ and (ii) the random choice in the tilt angle $\theta_0$ that cause decoherence over many experimental runs. We did so by deriving the dephasing rate induced by these noise channels, under a statistical model for the fluctuations. The analysis was carried out in a Lagrangian framework: rather than introducing noise phenomenologically by assuming the transfer function, we treat the noise in the acceleration and the tilt angle individually as stochastic inputs in the action, compute the resulting stochastic phase difference between the two interferometer arms, and then obtain the dephasing rate by implementing the Wiener-Khinchin theorem~\cite{Wiener:1930,Khintchine1934} for Gaussian noise statistics. This approach makes transparent (a) which physical terms in the dynamics are responsible for dephasing, and (b) how geometry and control parameters enter the susceptibility of the system to fluctuations. We then demonstrated the results for the case of white-noise statistics, i.e. constant power spectral densities for the acceleration and tilt fluctuations across frequency. 

The bounds on the PSD are obtained by imposing a target coherence threshold $\Gamma \tau\leq 1$, where $\Gamma$ is the dephasing rate and $\tau$ is the completion time for the one-loop interferometer in our particular setup, by fixing the values of system parameters according to 
($\eta_0=6\times 10^{3}{\rm \,T\,m^{-1}}, m=10^{-15}{\rm kg},~\tau=2\pi/\omega_0\simeq0.015$s), which yields $\Delta x =1{\rm nm}$ of spatial superposition. We tabulate the bounds on the PSD in Table.\ref{tab:accel_noise_bounds}).  

\begin{table}[ht!]
\caption{We denote the bounds on the square root of the PSD for the parameters
$m=10^{-15}$ kg, $\Delta x=1$nm, and $\tau=0.015$s, where $a$ denotes the external acceleration to the interferometer and $\theta_0$ is the angle between the spatial superposition axis and the acceleration noise. We emphasize that the values of $a_m$ and $\theta_m$—which mark regimes in which the system is comparatively less susceptible to acceleration noise—depend on other experimental parameters; the values quoted here are provided for illustrative purposes. For the discussion on
$a_m$, see Fig.~\ref{fig:rtS_aa_vs_a_-02-02}, and $\theta_m$, see Fig.~\ref{fig:rtS_aa_vs_theta_9.8_89.6-90}. We note that the values of $a_m,~\theta_m$ are based on the choice of the parameters, and they may have higher order corrections, which we have not taken in our current analysis.}
\label{tab:accel_noise_bounds}
\centering
\begin{ruledtabular}
\begin{tabular}{c D{.}{.}{-1} c}
$\theta_0\ (\mathrm{deg})$
& \multicolumn{1}{c}{$a\ (\mathrm{m\,s^{-2}})$}
& $\sqrt{\mathcal{S}_{aa}}\ (\mathrm{m\,s^{-2}\,Hz^{-1/2}})$ \\
\hline
0       & \multicolumn{1}{c}{$a_m=0.03$} & $\mathcal{O}(10^{-8})$  \\
0       & 0    & $\mathcal{O}(10^{-11})$ \\
0       & 3    & $\mathcal{O}(10^{-13})$ \\
89      & 9.81 & $\mathcal{O}(10^{-10})$ \\
$\theta_m$ = 89.825  & 9.81 & $\mathcal{O}(10^{-6})$  \\
\end{tabular}
\end{ruledtabular}
\end{table}

{The relaxed bounds, such as $\mathcal{O}(10^{-8})$ and $\mathcal{O}(10^{-6})\,\mathrm{m\,s^{-2}\,Hz^{-1/2}}$, are within the scale of state-of-the-art acceleration sensing and vibration-isolation experiments \cite{Timberlake2019,Lewandowski2021,Monteiro2020,Monteiro2017}. By contrast, the stricter bounds,
are demanding requirements for a tabletop Stern--Gerlach nanodiamond interferometer. Acceleration noise below this level has been demonstrated in highly specialized systems, most notably space-based free-fall experiments \cite{Armano2016}, but such performance is not currently routine in tabletop matter-wave interferometry. We note that the mechanism of attaining these constraints and the frequency range in which the PSD constraint is required will change between different setups, and is beyond the scope of this paper.}

%%%%%%%%%%%%%%%%%%%%%%%%
For $\theta_0 = 0^\circ$, the operating regime with minimum sensitivity to noise is when $a_m$ = 0.03 $\mathrm{m\,s^{-2}}$. Similarly, when $a = 9.81\ \mathrm{m\,s^{-2}}$, the tilt angle for which the system has minimum sensitivity to external acceleration fluctuation is $\theta_m= 89.825^\circ$. However, note that upper bound on the acceleration noise parameter quickly becomes stricter in the vicinity of the minimum sensitivity regions.

We also computed the bound on the PSD of the tilt angle, when the gravitational acceleration ($a= 9.81 {\rm \,m\, s^{-2}}$) is acting perpendicular to the spatial superposition, and obtained the following bound to satisfy the coherence constraint of $\Gamma\tau <1$:
$\sqrt{\mathcal{S}_{\theta\theta}} \lesssim \order{10^{-10}}\, \text{rad}\,\text{Hz}^{-1/2}$.

{The tilt-noise bound is also stringent but comparable to the level reached by 
precision tiltmeter technology \cite{Allocca2021}. However, we again note that the mechanism of attaining these constraints and the frequency range in which the PSD constraint is required will change between different setups, and is beyond the scope of this paper.}

We showed that the bound on acceleration-noise $\sqrt{\mathcal{S}_{aa}}$ becomes progressively stricter as the axis of spatial superposition aligns with the direction of the external random acceleration. However, the bound on tilt-noise $\sqrt{\mathcal{S}_{\theta\theta}}$ decreases with the alignment along the acceleration direction, for a fixed dephasing rate. 

Finally, while the white-noise assumption provides a clean and conservative baseline for requirements, {realistic acceleration and gravity-gradient noise spectra are generally colored (frequency-dependent), 
rather than white, especially in seismic and Newtonian-noise environments 
\cite{Saulson:1984,Harms_2019,RelAcc,Hughes_1998}. The white-noise approximation should be viewed as an analytic benchmark rather than a universal worst-case model. It is conservative if the measured PSD over the range of frequencies considered is lower than the constant white PSD constraints discussed above.} The Lagrangian framework developed here is naturally extensible to such scenarios by replacing the flat PSD with measured or modelled spectra {in Eq.\eqref{eq.second_cor_acc} and Eq.\eqref{eq.deph_rate_thetafluc}} and by incorporating cross-correlators to characterise the noise source and, hence, estimate the dephasing.

\begin{acknowledgments} 
We would like to thank Tracy Northup for the discussions related to the acceleration noise in the experimental setup. 
A.M.’s, S.B.’s and A.G.’s research is funded by the
Gordon and Betty Moore Foundation through Grant
GBMF12328, DOI 10.37807/GBMF12328. This material
is based on work supported by the Alfred P. Sloan 
Foundation under Grant No. G-2023-21130. A.G. also 
acknowledges support from NSF grants PHY-2409472 and
PHY-2111544, DARPA, the John Templeton Founda-
tion, the W.M. Keck Foundation, and the Simons 
Foundation. SNM would like to thank the organizers of the workshop, Schrodinger Cats: The quest to find the edge of the quantum world, held at OIST, Japan for facilitating academic interactions that led to this collaboration. SNM acknowledges support from the Kishore Vaigyanik Protsahan Yojana (KVPY) fellowship, SX-2011055, awarded by the Department of Science and Technology, Government of India.
 SB would like
to acknowledge EPSRC grants (EP/N031105/1,
EP/S000267/1, and EP/X009467/1) and grant
ST/W006227/1. 

\end{acknowledgments}

\section*{Data Availability}
The data that support the findings of this article are openly available \cite{SnehaAccTiltData2026}.

\bibliography{References}
\begin{appendix}
\appendix
\section{Contribution of trajectory and velocity fluctuations to dephasing}\label{App.fluc contribution}
Consider a Lagrangian $\mathcal{L}(\dot{x}, x, t)$. The equation of motion is then given by:
\begin{equation}
    \frac{d}{dt}\frac{\partial \mathcal{L}}{\partial \dot{x}} - \frac{\partial \mathcal{L}}{\partial x} = 0 \label{eq.EOM_Lag_gen}
\end{equation}
Consider the fluctuation in the the Lagrangian, $\delta\mathcal{L}$ introduced by some noise source $\delta \zeta$.
\begin{equation}
    \delta \mathcal{L} = \frac{\partial \mathcal{L}}{\partial \dot{x}}\delta \dot x + \frac{\partial \mathcal{L}}{\partial x}\delta x + \frac{\partial \mathcal{L}}{\partial \zeta}\delta \zeta \label{eq.fluc_Lag_gen}
\end{equation}
Substituting for $\frac{\partial \mathcal{L}}{\partial x}$ from eq.\eqref{eq.EOM_Lag_gen} in eq.\eqref{eq.fluc_Lag_gen}, we obtain:
\begin{align}
    \delta \mathcal{L} &= \frac{\partial \mathcal{L}}{\partial \dot{x}}\delta \dot x + \frac{d}{dt}\frac{\partial \mathcal{L}}{\partial \dot{x}}\delta x + \frac{\partial \mathcal{L}}{\partial \zeta}\delta \zeta\nonumber\\
    &= \frac{d}{dt}\left(\frac{\partial \mathcal{L}}{\partial \dot{x}}\delta x\right) + \frac{\partial \mathcal{L}}{\partial \zeta}\delta \zeta\label{eq.MZ1}
\end{align}

Consider $\delta \phi$ arising from $\delta \mathcal{L}$:
\begin{align}
    \delta \phi = & \frac{1}{\hbar}\int_0^Tdt\,\delta\mathcal{L}(\dot{x}, x, t)\nonumber\\
    = & \frac{1}{\hbar}\int_0^Tdt\,\left[\frac{d}{dt}\left(\frac{\partial \mathcal{L}}{\partial \dot{x}}\delta x\right) + \frac{\partial \mathcal{L}}{\partial \zeta}\delta \zeta\right]\label{eq.MZ2}
\end{align}
where $T = \frac{2\pi}{\omega_0}$. $\delta x(t)$ is defined upto its statistical properties deriving from the statistical properties of $\delta \zeta$ and the modified equations of motion. However, for each experimental run, $\delta x(t)$ will be a continuous and differentiable function, since it is describing the classical trajectory of a particle. Hence, we can apply the fundamental theorem of calculus (FTC). Now imposing the FTC on the first term, we obtain:
\begin{align}
    \delta \phi = & \frac{1}{\hbar}\left(\frac{\partial \mathcal{L}}{\partial \dot{x}}\delta x\right)\bigg|_{t=0}^{t=T} + \frac{1}{\hbar}\int_0^Tdt\frac{\partial \mathcal{L}}{\partial \zeta}\delta \zeta \label{eq.phase_fluc_final}
\end{align}
Now we consider the following type of Lagrangian:
\begin{equation}
    \mathcal{L}(\dot x, x, t) = \frac{1}{2}A\dot x^2 - V(x,t)
\end{equation}
where $A$ is a constant, and $V(x,t)$ is a function of position and time, independent of velocity. Thus, 
\begin{equation}
    \frac{\partial \mathcal{L}}{\partial \dot{x}}\delta x = A\dot x \delta x
\end{equation}
In a one loop interferometer, $\dot x(T) = 0 = \dot x(0)$. Thus, the first term in eq.\eqref{eq.phase_fluc_final} vanishes. This tells us that the phase fluctuation arises only from the direct fluctuation of acceleration and not from the induced trajectory and the velocity perturbations.

\section{Remarks on correlation statistics}\label{App.cor_stat}
For processes $A(t),B(t)$ define the cross-correlation by
\begin{equation}
R_{AB}(\tau)\;\equiv\;\mathbb{E}\!\big[A(t+\tau)\,B^{*}(t)\big],
\end{equation}
and the corresponding (cross-)power spectral density (PSD) by the Fourier transform
\begin{equation}
S_{AB}(\omega)\;\equiv\;\int_{-\infty}^{\infty} d\tau\, e^{i\omega\tau}\,R_{AB}(\tau).
\end{equation}

Let
\begin{equation}
Z(t)=X(t)-Y(t).
\end{equation}
Then the autocorrelation of $Z$ is
\begin{align}
R_{ZZ}(\tau)
&=\mathbb{E}\!\big[(X(t+\tau)-Y(t+\tau))(X^{*}(t)-Y^{*}(t))\big] \nonumber\\
&=\mathbb{E}\!\big[X(t+\tau)X^{*}(t)\big]+\mathbb{E}\!\big[Y(t+\tau)Y^{*}(t)\big]\nonumber\\
&\quad-\mathbb{E}\!\big[X(t+\tau)Y^{*}(t)\big]-\mathbb{E}\!\big[Y(t+\tau)X^{*}(t)\big]\nonumber\\
&=R_{XX}(\tau)+R_{YY}(\tau)-R_{XY}(\tau)-R_{YX}(\tau).
\end{align}

Fourier transforming term-by-term gives
\begin{equation}
S_{ZZ}(\omega)=S_{XX}(\omega)+S_{YY}(\omega)-S_{XY}(\omega)-S_{YX}(\omega).
\end{equation}

For wide sense stationary processes, the kind that we assume in this paper, one has $S_{YX}(\omega)=S_{XY}^{*}(\omega)$ . Hence
\begin{equation}
S_{ZZ}(\omega)=S_{XX}(\omega)+S_{YY}(\omega)-2\,\Re\!\left\{S_{XY}(\omega)\right\}. \label{eq.PSD_ZZ}
\end{equation}

If X and Y are uncorrelated if and only if $\Re\{S_{XY}(\omega)\} = 0$. Then the effective power spectral density is just the some of the PSD of each, X and Y. If $\Re\{S_{XY}(\omega)\}<0$, then the fluctuations between X and Y are anti-correlated. This increases the effective power spectral density more than the uncorrelated case. If $\Re\{S_{XY}(\omega)\}>0$, then the two variables are correlated, and the effective PSD is lesser than the uncorrelated case. Note that the autocorrelation power spectral density ($S_{XX}(\omega)$ \& $S_{YY}(\omega)$) is positive semi-definite.

In case of an interferometer dephasing, we consider the differential phase fluctuations (i.e., $\delta\phi = \frac{1}{\hbar}\int_0^Tdt\;(L_R-L_L) \equiv \delta\phi_R - \delta\phi_L$. Hence, the effective PSD can be modelled as given in eq.\eqref{eq.PSD_ZZ}. We argue that in case of the noise correlation between the arms of a Stern-Gerlach type interferometer, is a positive correlation (Refer footnote(8)). Thus, computing the dephasing rate under the assumption of uncorrelated fluctuations imposes an upper bound on it. Similar conclusions were brought in a slightly different context in~\cite{Wu:2024bzd}.

\newpage{
\section{Verifying the smoothness of plots}\label{App.zoom}
Fig.\ref{fig:rtS_aa_vs_a_zoom} depicts the smoothness of the plot Fig.\ref{fig:rtS_aa_vs_a_-02-02} in the regime where the interferometer is less sensitive to acceleration noise.
\begin{figure}[ht!]
    \centering
    \includegraphics[width=\linewidth]{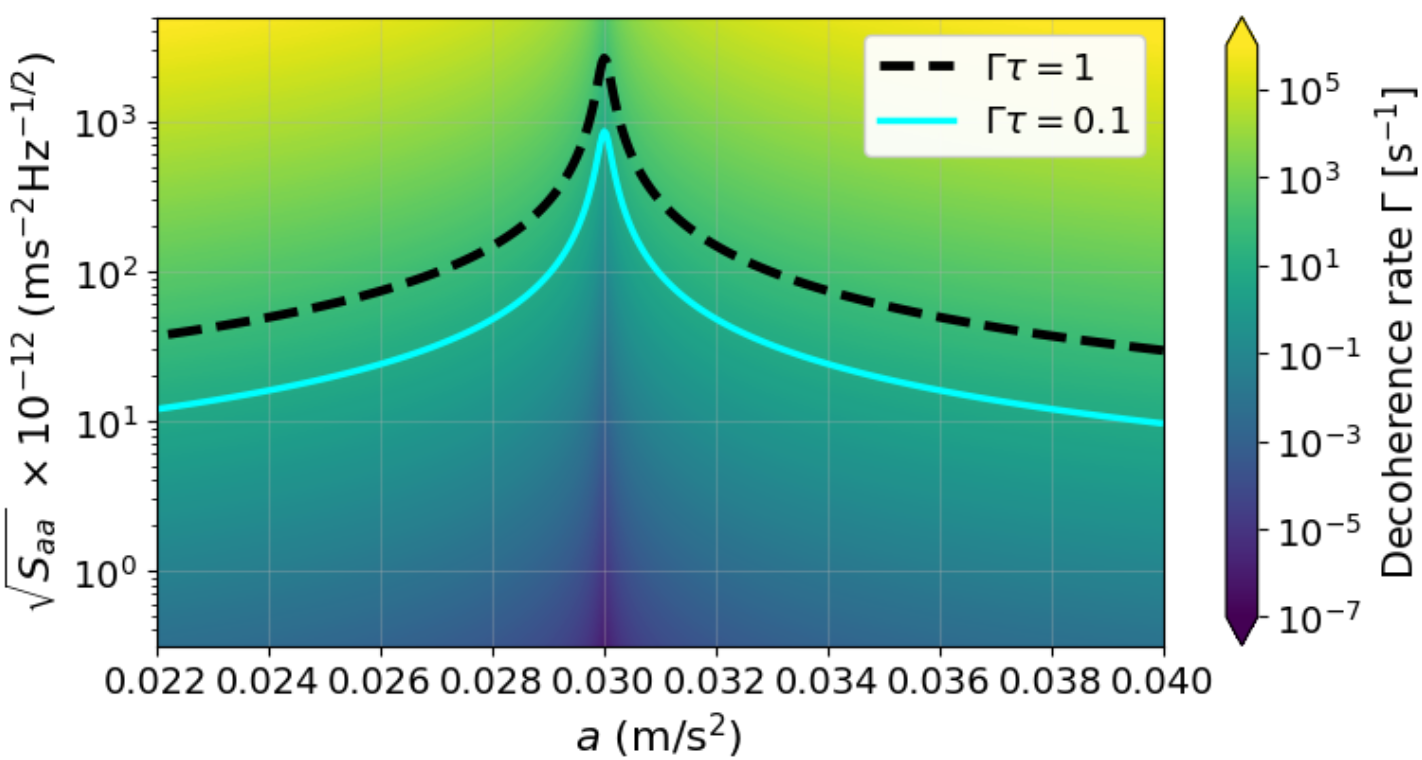}
    \caption{This plot showcases the of the plot Fig.\ref{fig:rtS_aa_vs_a_-02-02} in the parameter regime where the interferometer has relatively higher tolerance to acceleration fluctuations.}
    \label{fig:rtS_aa_vs_a_zoom}
\end{figure}

Fig.\ref{fig:rtS_aa_vs_theta_9.8_zoom} depicts the smoothness of the plot Fig.\ref{fig:rtS_aa_vs_theta_9.8_89.6-90} in the regime where the interferometer is less sensitive to acceleration noise.
\begin{figure}[ht!]
    \centering
    \includegraphics[width=\linewidth]{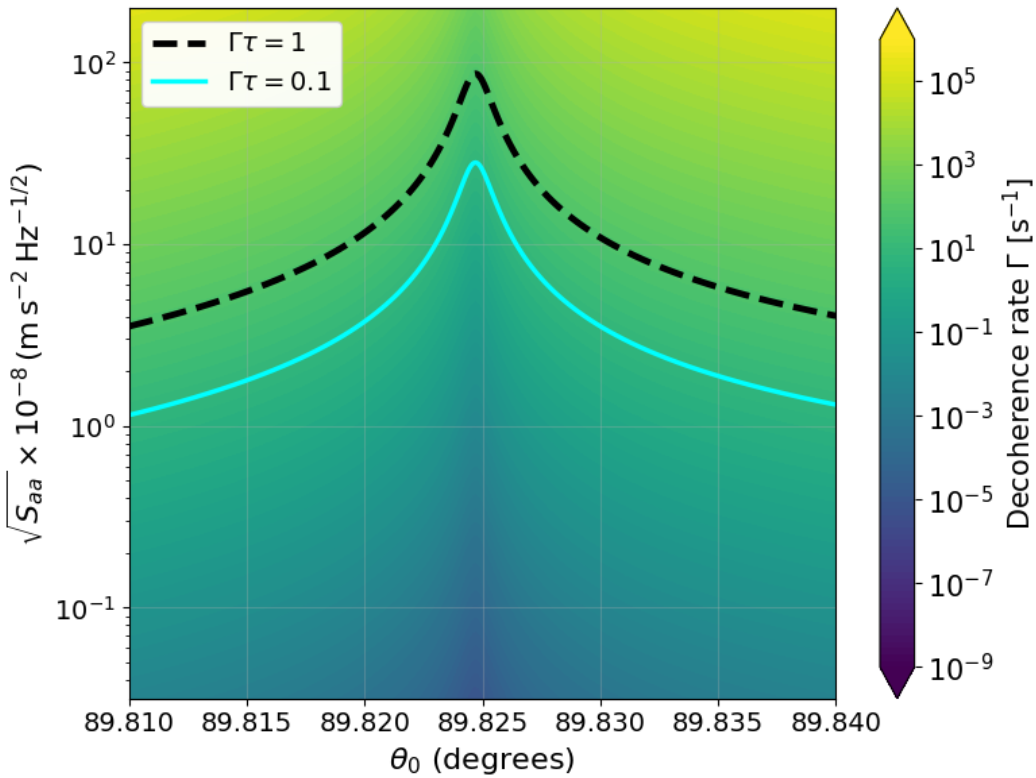}
    \caption{This plot showcases the of the plot Fig.\ref{fig:rtS_aa_vs_theta_9.8_89.6-90} in the parameter regime where the interferometer has relatively higher tolerance to acceleration fluctuations.}
    \label{fig:rtS_aa_vs_theta_9.8_zoom}
\end{figure}
}

%\section{Noise to signal ratio bounds}In this section, we compute the tolerable noise to signal ratio for the gravitational acceleration and tilt angle noise using the bounds obtained in the text. We define the noise amplitude as $\sqrt{\omega_0} A_a$ for acceleration noise and $\sqrt{\omega_0} A_\theta$ for tilt angle noise\footnote{From eq.\eqref{eq.whitenoise_stat2} in the time-domain, we can consider $\sqrt{2\pi} A_a$ to be proportional to the noise amplitude. By dimensional analysis we see that we need a system-dependent frequency term, which in our case is $\omega_0/2\pi$. Hence, the noise amplitude is motivated to be defined as $\sqrt{\frac{\omega_0}{2\pi}}\sqrt{2\pi} A_a = \sqrt{\omega_0} A_a $}. 
\end{appendix}
\end{document}